\documentclass[11pt,a4paper]{article}

\usepackage{jcappub}

\DeclareMathAlphabet{\mathpzc}{OT1}{pzc}{m}{it}

\newcommand{\ws}{\textcolor{white}{!}}
\newcommand{\be}{\begin{equation}}
\newcommand{\ee}{\end{equation}}
\newcommand{\bes}{\begin{equation*}}
\newcommand{\ees}{\end{equation*}}
\newcommand{\fett}[1]{\boldsymbol{#1}}
\newcommand{\al}{\alpha(\eta)}
\newcommand{\dd}{{\rm{d}}}
\newcommand{\sign}{\textcolor{white}{-}}

\newcommand{\myTh}[2]{\vskip #1 cm\hspace{#2 cm} ($i$ = 1,2,3) \vskip -#1 cm \vskip-1.2cm
\noindent} 

\newcommand{\ii}{{\rm i}}

\makeatletter
\def\fleq{\@fleqntrue\let\mathindent\@mathmargin \@mathmargin=\leftmargini}
\def\cneq{\@fleqnfalse}
\makeatother

\gdef\@fpheader{blubb}

\begin{document}

\begin{flushright}
{\large \tt 
TTK-12-09}
\end{flushright}

\title{Lagrangian perturbations and the matter bispectrum I: fourth-order model for non-linear clustering}

\author[a]{Cornelius~Rampf$\,$}
\author[b]{and Thomas Buchert$\,$} 

\affiliation[a]{Institut f\"ur Theoretische Teilchenphysik und Kosmologie, 
  RWTH Aachen, Physikzentrum RWTH--Melaten, D--52056 Aachen, Germany}
\affiliation[b]{Universit\'e de Lyon, Observatoire de Lyon,
Centre de Recherche Astrophysique de Lyon, CNRS UMR 5574: Universit\'e Lyon~1 and \'Ecole Normale Sup\'erieure de Lyon, \\
9 avenue Charles Andr\'e, F--69230 Saint--Genis--Laval, France}

\emailAdd{rampf@physik.rwth-aachen.de, buchert@obs.univ-lyon1.fr}

\abstract{
We investigate the Lagrangian perturbation theory of a homogeneous and isotropic universe in the non-relativistic limit, and derive the solutions up to the fourth order.
These solutions are needed for example for the next-to-leading order correction of the (resummed) 
Lagrangian matter bispectrum, which we study in an accompanying paper. 
We focus on flat cosmologies with a vanishing cosmological constant, 
and provide an in-depth description of two complementary approaches used in the current literature. 
Both approaches are solved with two different sets of initial conditions---both appropriate for 
modelling the large-scale structure. Afterwards we consider only the fastest growing mode solution, which is not affected by either of these choices of initial conditions.
Under the reasonable approximation that the linear density contrast is evaluated at the initial Lagrangian position of the fluid particle, we obtain the $n$th-order displacement field in the so-called \emph{initial position limit}: the $n$th order displacement field consists of 3(n-1) integrals over n linear density contrasts, and obeys self-similarity. 
Then, we find exact relations between the series in Lagrangian and Eulerian perturbation theory, leading to identical predictions for the density contrast and the peculiar-velocity divergence up to the fourth order.}

\maketitle

\flushbottom

%%%%%%%%%%%%%%%%%%%%%%%%%%%%%%%%%%%%%%%%%%%%%%%%%%%%%%%%%%%%%%%%%%%
\section{Introduction}

In the last years several analytic techniques have been proposed in order to study the inhomogeneities of the 
large scale structure (LSS)
\cite{Taylor:1996ne,Buchert:1992ya,Bouchet:1992uh,Bouchet:1994xp,Scoccimarro:2000zr,McDonald:2006hf,Matsubara:2007wj}. 
The basic idea of them is to solve the equations for an irrotational and pressureless fluid of cold dark 
matter particles in terms of a perturbative expansion. In the standard scenario the density and velocity field
of the fluid particle are the perturbed quantities. Thus the validity of the perturbative series depends on the smallness 
of these fields. This approach is called Eulerian (or standard) perturbation theory (SPT), since the equations are
evaluated as a function of Eulerian coordinates \cite{Bernardeau:2001qr}. Subsequent gravitational collapse leads to
highly non-linear structures in the universe like galaxies, clusters of galaxies, etc., i.e., regions where the local density 
field departs significantly from the mean density. As a result, the series in SPT breaks down.
This situation was already realised in 1969 by Zel'dovich, who proposed an approximate solution which is above all applicable
to the highly non-linear regime by a Lagrangian extrapolation of the Eulerian linear solution, inspired by the exact solution for inertial systems \cite{zeldovich:fragmentation,zeldovich:fragmentation2,zeldovich:rev,zeldovichmyshkis}: the Zel'dovich approximation (ZA). 
In general, the ZA can be derived from the full system of gravitational equations and forms a subclass of solutions of the Lagrangian theory of gravitational 
instability (i.e., Lagrangian perturbation theory; LPT) \cite{Buchert:1989xx,Buchert:1992ya,Buchert:1993mp,Buchert:1993xz,Buchert:1993ud}. In LPT 
the only perturbed quantity is the \emph{gravitational induced deviation} of the particle trajectory field from the homogeneous background
expansion. Stated in another way, LPT does not rely on the smallness of the density and velocity fields, but on the smallness of the 
deviation of the trajectory field, in a coordinate system that moves with the fluid. 
It can be shown that this implies a weaker constraint on the validity of the series and hence can be maintained substantially longer during the gravitational
evolution (see the thorough discussion in \cite{chernin,Munshi:1994zb,Sahni:1995rm}). Additionally, to obtain an SPT series one basically has to approximate the continuity- and the Euler-equation order by order, 
so that, strictly speaking, mass- and momentum-conservation are not fulfilled. In LPT, on the other hand, the Jacobian
of the transformation from the Eulerian to the Lagrangian frame is approximated and so is the precise localisation of the fluid element,
whereas the continuity- and Euler-equation are still exactly solved.

General perturbation and solution schemes to any order of Lagrangian perturbations on any FLRW background have been given in the review \cite{ehlersbuchert}, based on explicit 
evaluations of the general first-order scheme including rotational flows \cite{Buchert:1989xx,Buchert:1992ya,bildhauer:1992}, and
the general second-order scheme for irrotational flows \cite{Buchert:1993xz}. 
The third-order scheme for irrotational flows with \emph{slaved} initial conditions (i.e., for an assumed initial parallelism of peculiar-velocity and peculiar-acceleration)  is given in \cite{Buchert:1993ud}, and 
the fourth-order scheme for this subclass of the general solution has been derived in \cite{vanselow} (see further below for an explanation). 
Lagrangian perturbation theory has also been extended to include 
pressure \cite{adlerbuchert}, and the series can be derived from exact integrals of longitudinal and transverse parts \cite{buchert:integral,buchert:nonperturbative}. Extensive comparisons of LPT results against N-body simulations can be found in \cite{Buchert:1995km,Karakatsanis:1996uz,Weiss:1995xk,Melott:1994ah,Bouchet:1994xp,Buchert:1993df,Shandarin3,Pauls:1994pp,chernin}.

In the present paper, we explicitly reexamine the Lagrangian framework in two different representations and evaluate them up to the fourth order. For both representations we choose a different set of initial conditions, which can be labeled as `Zel'dovich type', since only one initial potential has to be chosen instead of two in the general case. Note that the fastest growing mode solution is not affected by either of these choices.
At this point the reader may ask, what is the point in deriving higher-order solutions for the purpose of modelling the LSS. First of 
all, the fourth-order solution is needed for the next-to-leading order correction to  the LPT matter bispectrum, which we calculate in an accompanying
paper. From a theoretical point of view one also expects a match between SPT and LPT under certain circumstances. In reference \cite{Matsubara:2007wj}
the equivalence of SPT and LPT is shown, if one sums up the perturbative solutions up to the third order.
However it is not clear \emph{a posteriori} if this matching between SPT and LPT occurs at the fourth-order as well, because the convergence of
the LPT and SPT series need not to behave equally. Furthermore, there is a growing interest to apply 
higher-order solutions in the context of resummation approaches, and we consider an explicit demonstration and a thorough comparison with the SPT series useful. Also, 
note that resummation techniques in the LPT framework are directly feasible rather than complex scenarios in SPT \cite{RampfWong:2012}.

Although the subject of this paper is quite technical, we try to keep it as readable as possible, e.g., we restrict our calculations to an
Einstein-de Sitter (EdS) universe. The organisation of this paper is as follows: in section \ref{secSys} we derive the evolution equations step by
step and \textit{confront} two complementary ways of how to deal explicitly with LPT calculations. We do so to shed light on two different looking
formalisms used in the current literature. Then, in section \ref{secIC} we mention and explain our choice of initial conditions.
In section \ref{secSol1} and \ref{secSol2} we show the results in both formalisms. Afterwards, we prepare our solutions to be used in Fourier space and derive relations between the SPT and LPT series in section \ref{sec:realisation}. Finally, in section \ref{sec:discussion} we give a discussion and conclude. Our notation is introduced and defined in the text, but is also summarised in table \ref{tab:notation}.

%%%%%%%%%%%%%%%%%%%%%%%%%%%%%%%%%%%%%%%%%%%%%%%%%%%%%%%%%%%%%%%%%%
\section{Systems of equations}\label{secSys}

\hypersetup{linkcolor=black}

According to the $\Lambda$CDM model and its current success in treating 
most of the problems in observational cosmology \cite{Larson:2010gs},
we live in a statistically homogeneous and isotropic universe. 
The universe is expanding, thus the mean density $\overline{\rho}(t)$ 
is diluting. But this global effect cannot compete with the gravitational potential
\textit{locally}. Hence local density fluctuations $\delta(\fett{r},t)$ 
are the source of gravitational collapse, which leads to the observed LSS.
We can define the above quantities in terms of the full density $\rho(\fett{r},t)$
as
\be \label{density}
 \rho(\fett{r},t) = \overline{\rho}(t) [1+ \delta(\fett{r},t)] \,,
\ee
where $\overline{\rho}$ is given by the assumed homogeneous background density in a Newtonian model with hypertorus topology
\cite{buchertehlers}.
In this chapter we set up the evolution equations that are sourced by 
\begin{itemize}
 \item \textbf{\ref{secSys}.n.1}: the full density $\rho(\fett{r},t)$, 
  which we label with ``full-system approach'',
 \item \textbf{\ref{secSys}.n.2}: and for the density contrast $\delta(\fett{r},t)$, the ``peculiar-system approach'',
\end{itemize}
step by step (\textbf{n} = \textbf{1}, $\cdots$, \textbf{4}) and independent from each other. 
\hypersetup{linkcolor=blue}
Readers who are only interested in the final equations may go directly to page \pageref{LN1fin}: 
Eqs.\,(\ref{LN1fin}-\ref{LN3fin}) are the equations in the full-system approach, whereas
Eqs.\,(\ref{LN4fin}-\ref{LN6fin}) refer to the peculiar-system approach.
The perturbation equations are shown in section \ref{secPert}. Finally, in section \ref{secComp}, we clarify errors 
and common misunderstandings in the current literature.

\subsection{Eulerian equations}

\subsubsection{Full-system approach}

Let us briefly go through the derivation of the equations of motion in the Lagrangian description. 
For simplicity we focus on flat cosmologies with a vanishing cosmological constant although a more general
implementation is straightforward \cite{Buchert:1989xx, Buchert:1992ya, Buchert:1995bq}. 
As usual we denote the density, the velocity and the acceleration fields by $\rho$, $\fett{v}$, and
$\fett{g}$, respectively. In a non-rotating (Eulerian) frame with  coordinate
$\fett{r}$ at cosmic time $t$, the equations for self-gravitating and irrotational dust are \cite{Buchert:1987xy}:
\begin{align} 
  \label{conti}       \frac{\partial \rho(\fett{r},t)}{\partial t} 
            + \fett{\nabla_r} \cdot &\left[\rho(\fett{r},t)\, \fett{v}(\fett{r},t) \right] = 0  \,, \\
  \label{velocity}  \varepsilon_{ijk} \, \partial_{r_j} v_k (\fett{r},t) &= 0 \,,  \\
  \label{irrot}     \varepsilon_{ijk} \, \partial_{r_j} g_k (\fett{r},t) &= 0 \,,  \\
  \label{Euler}     \fett{\nabla_r} \cdot \dot{\fett{v}}(\fett{r},t) 
            &= -4\pi G \rho (\fett{r},t) \, , \,\,\,\,\, {\rm{with}} \,\,\,
      \dd \fett{v}/\dd t \equiv \dot{\fett{v}} =   \fett{g}  \,\,{\rm{and}}  \,\,\, \rho >0 \,,
\end{align}
\myTh{-1.64}{10.4}

\vspace{0.1cm}

\noindent where Einstein summation over the spatial (Eulerian) components is implied. Eq.\,(\ref{conti}) is the continuity equation and denotes 
mass conservation, Eq.\,(\ref{velocity}) states the irrotationality of the velocity, and should be viewed
as an additional constraint to the field equation that requires an irrotational acceleration field, i.e., to Eq.\,(\ref{irrot}). 
The divergence of the field strength, here with Euler's equation inserted, is linked to the density source in Eq.\,(\ref{Euler}).
Note that we make use of the convective time derivative, i.e., 
$\dd / \dd t := \left. \partial / \partial t  \right|_{\fett{r}} + \fett{v} \cdot \fett{\nabla_r}$,
which we denote by an overdot.

\subsubsection{Peculiar-system approach}

Eqs.\,(\ref{conti})-(\ref{Euler}) are written in terms of the full density $\rho(\fett{r},t)$, 
thus including the homogeneous and isotropic deformation of an expanding universe $\overline \rho(t)$ 
($\equiv \overline{\rho}_0/a^3$) for a matter dominated universe). 
However, it is also possible to construct a set of equations 
\cite{Buchert:1989xx,Bouchet:1992uh,Catelan:1994ze}
where Poisson's equation is only sourced by the density contrast $\delta$, which is linked to the 
full density in the following way: $\rho(\fett{x},t) = \overline \rho(t) \left[ 1+ \delta(\fett{x},t) \right]$.
$\fett{x}$ denotes the comoving distance and is related to the physical distance as
$\fett{r} = a \,\fett{x}$, where $a(t)$ is the cosmic scale factor.
The Poisson equation then reads \cite{Peebles:1980,Shandarin1,Shandarin2,Buchert:1989xx}:
\be
 \label{pois} \Delta_{\fett{x}} \Phi(\fett{x},\eta) = \alpha(\eta) \,\delta (\fett{x},\eta) \,, \,\,\,\, \alpha = 6/(\eta^2 +k )\,,
\ee
where we have switched to superconformal time $\eta \equiv \sqrt{-k} (1-\Omega)^{-1/2}$ (we denote 
its derivatives with $\dd / \dd \eta \equiv\, '$) \cite{Catelan:1994ze}. For an EdS universe we have the simplification $a^2\dd \eta = \dd t$, with $a = (\eta_0/\eta)^2$, 
and $\alpha = 6/\eta^2$.
The peculiar-evolution equations can then be written as (compare with Eqs.\,(\ref{conti}-\ref{Euler})):
\vspace{0.2cm}
\myTh{1.37}{10.5}
\begin{align}
 \label{contiSuper} &\frac{\partial \delta(\fett{x},\eta)}{\partial \eta} + \fett{\nabla_x} \cdot \left\{ \left[ 1+\delta(\fett{x},\eta) \right] \fett{u}(\fett{x},\eta)\,a \right\} = 0 \,, \\
 \label{eins} &\varepsilon_{ijk} \, \partial_{x_j} u_k (\fett{x},\eta) = 0  \,, \\
 \label{zwei} &\varepsilon_{ijk} \, \partial_{x_j} u_k'(\fett{x},\eta) = 0 \,, \\
 \label{drei} &\fett{\nabla_x} \cdot {\fett{g}}_{\rm pec}(\fett{x},\eta)\hspace{0.05cm}  = -\al \,\delta(\fett{x},\eta)   \, , \,\,\,\,\, {\rm{with}} \,\,\,
      \dd \fett{u}/\dd \eta \equiv \fett{u}' =   \fett{g}_{\rm pec}  \,\,{\rm{and}}  \,\,\, \delta > -1 \,,
\end{align}
Eq.\,(\ref{eins}) states the irrotationality of the peculiar-velocity $\fett{u}= a \,\dd \fett{x}/ \dd t $, 
Eq.\,(\ref{zwei}) the irrotationality of the (rescaled) peculiar-acceleration 
$\fett{g}_{\rm pec} \!\equiv\! \dd^2 \fett{x} / \dd^2 \eta$, and Eq.\,(\ref{drei}) links the acceleration 
to the density contrast field.

%%%%%%%%%%%%%%%%%%%%%%%%%%%%%%%%%%%%%%%%%%%%%%%%%%%%%%%%%%%%%%%%%%%
\subsection{From the Eulerian to the Lagrangian framework}

The two set of equations, namely Eqs.\,(\ref{velocity}-\ref{Euler}) and Eqs.\,(\ref{eins}-\ref{drei})
are equivalent but have technical subtleties with their pros and cons, which we point 
out in the following Lagrangian description. 
We now briefly recall the corresponding Lagrangian systems that have been
introduced in \cite{Buchert:1987xy} and \cite{Buchert:1989xx} in the full-system approach, and in \cite{Buchert:1989xx}, appendix A, for the peculiar-system.
Note that the Lagrangian description can be formulated in the same Lagrangian coordinates in both systems, which is the reason why the
peculiar-system is essentially redundant. We nevertheless have chosen to confront the two approaches for the purpose of assisting work that deals with either
of the two representations. Additionally, the peculiar-system approach is useful in order to link the LPT series to its counterpart in SPT, which we do so in section \ref{sec:realisation}.

\subsubsection{Full-system approach}

It is useful to transform from Eulerian coordinates $r_i$ to Lagrangian coordinates $q_i$ ($i\!=\!1,2,3$), and
we start with the transformation of Eqs.\,(\ref{conti}-\ref{Euler}). We 
introduce integral curves $\fett{r} = \fett{f}(\fett{q},t)$ of the velocity field
\be
 \label{curves} \frac{\dd \fett{f}}{\dd t} = \fett{v} \,, 
\ee
and set the initial position at time $t_0$ to
\be
  \label{q} \fett{f}(\fett{q},t_0) =: \fett{q} \,.
\ee
The Jacobian of the transformation can be written as
\be \label{Jacob}
  J_{ij} (\fett{q},t) := f_{i,j} \,, \qquad J:= \det[J_{ij}] \,,
\ee
where commas denote partial derivatives with respect to Lagrangian coordinates $\fett{q}$.
The formal requirement $J > 0$ guarantees the existence of regular solutions and the mathematical equivalence to the Eulerian system \cite{ehlersbuchert}.
The continuity equation, Eq.\,(\ref{conti}), is then integrated to yield $\rho = \rho (\fett{q},t_0)/J$, 
and with the usage\footnote{We assume
the equivalence of the acceleration field and the gravitational field strength, i.e., Einstein's equivalence principle.} of $\fett{g} =  \ddot{\fett{f}}$ we cast
Eqs.\,(\ref{velocity}-\ref{Euler}) into \cite{Buchert:1987xy}
\begin{align}
 \label{dot}  \varepsilon_{ijk} J_{lj}^{ -1} \dot  J_{kl} &= 0 \,,  \\
 \label{ddot} \varepsilon_{ijk} J_{lj}^{ -1} \ddot J_{kl} &= 0 \,, \\
\label{rho0}  J_{ij}^{ -1} \ddot J_{ji} &= - 4\pi G \rho_0 J^{-1} \,,\qquad\rho_0 = \rho (\fett{q},t_0) \,.
\end{align}
Note that Eq.\,(\ref{ddot}) follows directly from Eq.\,(\ref{dot}). Hence from the irrotationality 
of the velocity field we can conclude the irrotationality of the acceleration field 
(but not vice versa!). This is shown in \cite{Buchert:1993xz} and is valid for \textit{any} integral curves, 
and for the perturbative solutions at each order as well.
With the inverse Jacobian, i.e.,  $J_{ij}^{-1} \equiv 1/J \, {\rm{adj}} [ J_{ij} ]
= 1/ (2J)\, \varepsilon_{ilm} \varepsilon_{jpq} J_{pl} J_{qm}$, and the use of Eq.\,(\ref{Jacob}) we have:
\begin{align}
 \label{LN1} f_{i,n}\, \varepsilon_{njk} f_{l,j} \dot{f}_{l,k} &= 0 \,,  \\
 \label{LN2} f_{i,n}\, \varepsilon_{njk} f_{l,j} \ddot{f}_{l,k} &= 0 \,, \\
 \label{LN3} \varepsilon_{ilm} \varepsilon_{jpq} \ddot{f}_{j,i} f_{p,l} f_{q,m} &= - 8\pi G \rho_0 \,.
\end{align}
To solve Eqs.\,(\ref{LN1}-\ref{LN3}) we impose the following ansatz for the trajectory
\be \label{An1}
  \fett{f}(\fett{q},t) = a(t)\, \fett{q} + \fett{p}(\fett{q},t) \,\, \Rightarrow \,\, f_{i,j} = a \delta_{ij}+ p_{i,j} \,,
\ee
where $a \fett{q}$ stands for the homogenous-isotropic background deformation, and $\fett{p}$ is the perturba-tion---induced by
gravitational interaction. Plugging Eq.\,(\ref{An1}) into Eqs.\,(\ref{LN1}-\ref{LN3}) we finally obtain:
\begin{eqnarray}
 \label{LN1fin} (a^2 \frac{\dd }{\dd t}\:\,- \dot a a ) \,  \varepsilon_{ijk}\, p_{k,j} 
    =  a\, \varepsilon_{ijk}\,  p_{l,j} \,\dot{p}_{l,k} 
    + p_{i,n} \,\varepsilon_{njk} \left[p_{l,j}\, \dot{p}_{l,k} - (a \frac{\dd }{\dd t}\:\,- \dot a ) \,p_{k,j} \right] && \,, \\
 \label{LN2fin} (a^2 \frac{\dd^2 }{\dd t^2}- \ddot a a ) \,  \varepsilon_{ijk}\, p_{k,j} 
    =  a\, \varepsilon_{ijk}\,  p_{l,j} \, \ddot{p}_{l,k} 
    + p_{i,n} \,\varepsilon_{njk} \left[p_{l,j} \,\ddot{p}_{l,k} - (a \frac{\dd^2 }{\dd t^2}- \ddot a ) \,p_{k,j} \right] && \,, \\
 \label{LN3fin} (a^2 \frac{\dd^2}{\dd t^2} +2\ddot a a )\,p_{l,l} +a \,(p_{i,i}\, \ddot{p}_{j,j}- p_{i,j}\, \ddot{p}_{j,i})
   + \frac{\ddot a}{2} (p_{i,i}\, {p}_{j,j}- p_{i,j} \,{p}_{j,i}) +p_{i,j}^c\, \ddot{p}_{j,i} \quad&\ws& \nonumber \\ 
 \ws \hspace{9cm} = - (4\pi G \rho_0 +3 \ddot a a^2) &&\,.
\end{eqnarray}
In the above equations we have defined the co-factor element \mbox{$p_{i,j}^c\equiv\,1/2\,\varepsilon_{ilm} \varepsilon_{jpq} p_{p,l}\, p_{q,m}$}.
Eqs.\,(\ref{LN1fin}-\ref{LN3fin}) with $\rho_0 \!>\! 0$ form a closed set of Lagrangian evolution equations in the full-system approach.

\subsubsection{Peculiar-system approach}\label{sec:pecSys}

Analogous considerations lead to a similar set for Eqs.\,(\ref{eins}-\ref{drei}): 
the comoving trajectory field is $\fett{x} = {\fett{F}}(\fett{q},\eta)$, and with $\fett{g}_{\rm pec} =  {\fett{F}}''$ it follows that:
\begin{align}
 \label{LN4} F_{i,n}\, \varepsilon_{njk} F_{l,j} F_{l,k}' &= 0 \,, \\
 \label{LN5} F_{i,n}\, \varepsilon_{njk} F_{l,j} F_{l,k}'' &= 0 \,,\\
 \label{LN6} \varepsilon_{ilm} \varepsilon_{jpq} F_{j,i}''F_{p,l} F_{q,m} &= - 2\, \alpha \,\delta J^F \,,
\end{align}
where $J^F \equiv {\rm det}[F_{i,j}]$, and, as before, a prime denotes a derivative with respect to superconformal time. 
For the set of Eqs.\,(\ref{LN4}-\ref{LN6}) we impose the ansatz:
\be \label{An2}
  {\fett{F}}(\fett{q},\eta) = \fett{q} + \fett{\Psi}(\fett{q},\eta) \,\,\Rightarrow \,\, F_{i,j} =  \delta_{ij}+ \Psi_{i,j} \,, 
\ee 
where the crucial difference in Eq.\,(\ref{An2}) with respect to Eq.\,(\ref{An1}) is the missing factor of $a$, and $\fett{F} = \fett{f}/a$ links the comoving to the physical trajectory field.
In the peculiar-system approach it is common to choose mass conservation by the following constraint:
\be \label{mc}
   \rho(\fett{x},\eta) \,\dd^3 x \equiv \overline{\rho}(\eta)\, \dd^3 q \,, \qquad  \delta = 1/ J^F -1 \,.
\ee
Explicitly, in doing so we either restrict ourselves to a specific class of initial 
conditions \cite{Buchert:1992ya} or assumed initial quasi-homogeneity. We shall discuss this issue later in 
section \ref{secSol2}.

Plugging Eq.\,(\ref{An2}) into Eqs.\,(\ref{LN4}-\ref{LN6}) we have:
\begin{align}
 \label{LN4fin} \varepsilon_{ijk} \Psi_{k,j}' &= \varepsilon_{ijk}  \Psi_{l,j} \Psi_{l,k}' 
      + \Psi_{i,n} \,\varepsilon_{njk} \left(\Psi_{l,j} \Psi_{l,k}' - \Psi_{k,j}' \right) \,, \\
 \label{LN5fin} \varepsilon_{ijk} \Psi_{k,j}'' &= \varepsilon_{ijk}  \Psi_{l,j} \Psi_{l,k}''
    + \Psi_{i,n}\, \varepsilon_{njk} \left(\Psi_{l,j} \Psi_{l,k}'' - \Psi_{k,j}'' \right) \,, \\
 \label{LN6fin} \al [ J^F-1] &= \left[ (1+ \Psi_{l,l}) \delta_{ij} - \Psi_{i,j} 
   + \Psi_{i,j}^c  \right] {\Psi}_{j,i}'' \,.
\end{align}
Similar to above we have defined $\Psi_{i,j}^c \equiv\,1/2\,\varepsilon_{ilm} \varepsilon_{jpq} \Psi_{p,l}\, \Psi_{q,m}$.
Eqs.\,(\ref{LN4fin}-\ref{LN6fin}) with $J^F > 0$ is the closed set of Lagrangian evolution equations in the peculiar-system
approach.

%%%%%%%%%%%%%%%%%%%%%%%%%%%%%%%%%%%%%%%%%%%%%%%%%%%%%%%%%%%%%%%%
\subsection{The perturbation ansatz}

The full- and peculiar-systems are highly non-linear, thus it is common to seek approximate solutions in terms of perturbative series. 

\subsubsection{Full-system approach}

First we proceed with the ansatz for the full-system, i.e., Eqs.\,(\ref{LN1fin}-\ref{LN3fin}). 
We expand the perturbation $\fett{p}(\fett{q},t)$ into a series, and factorise
out the spatial and temporal dependence. Additionally, 
we decompose the  inhomogeneous 
deformation $\fett{p}(\fett{q},t)$ of the $n$th order into purely longitudinal contributions, 
which we denote by $\fett{\Psi}^{(n)}(\fett{q}) \equiv \fett{\nabla_q} \phi^{(n)}(\fett{q})$, 
and purely transverse contributions,\footnote{The Lagrangian transverse parts are mandatory in the Lagrangian frame to 
guarantee irrotationality in the Eulerian frame.} denoted by 
$\fett{T}^{(n)}(\fett{q}) \equiv \fett{\nabla_q} \times \fett{A}^{(n)}(\fett{q})$;
the temporal parts of the $n$th-order correction are denoted by $q_n(t)$ or $q_n^T(t)$:
\begin{eqnarray} \label{pert1}
  \fett{p}(\fett{q},t) = \epsilon\, q_1(t) \,\fett{\Psi}^{(1)}(\fett{q}) +\epsilon^2 q_2(t)\, \fett{\Psi}^{(2)}(\fett{q})
     &+& \epsilon^3 q_3(t) \,\fett{\Psi}^{(3)} (\fett{q})+ \epsilon^4 q_4(t) \, \fett{\Psi}^{(4)}(\fett{q}) \nonumber \\
   &+& \epsilon^3 q_3^T(t) \, \fett{T}^{(3)}(\fett{q})+ \epsilon^4 q_4^T(t) \,\fett{T}^{(4)}(\fett{q}) \,.
\end{eqnarray}
The parameter $\epsilon$ is supposed to be small and dimensionless. In the most general cases,\footnote{See the general second-order solution for irrotational flows given in \cite{Buchert:1993xz}.} 
nontrivial  solutions of the irrotationality condition
(i.e., $\fett{\nabla_q} \times \fett{T}^{(n)} \neq \fett{0}$) of Eq.\,(\ref{LN1}) and Eq.\,(\ref{LN2}) occur the first time 
at the order of $\epsilon^2$ 
and are henceforth required  for higher-order solutions.
It should be pointed out that in addition to longitudinal (i.e.,\;\;potential) modes, 
Lagrangian transverse modes also affect the growth of density perturbations. 
Additional to the above perturbation ansatz, we
set the initial density $\rho_0$ (on the RHS in Eq.\,(\ref{LN3})) to be
\be \label{iniRho}
 \rho(\fett{q},t_0) = \overline{\rho}({t}_0) \left[ 1+ \epsilon \delta (\fett{q},t_0) \right] \,,
\ee
without loss of generality, where $\overline{\rho}(t_0)$ and 
$\delta(\fett{q},t_0)$ denote the initial background density 
and the initial density contrast, respectively. 
In section  \ref{secSol1} we shall need the above equation in order to set up the initial data.

\subsubsection{Peculiar-system approach}

The perturbative treatment of $\fett{\Psi}$ is quite analogous to the above. However,
because of the LHS in Eq.\,(\ref{LN6fin}), we need 
the explicit expansion of the Jacobian as well. 
This is clearly a (technical) disadvantage in comparison with the evolution equations in the full-system approach, where the Jacobian cancels out (but only in the absence of a cosmological constant). 

The Jacobian of the transformation between the 
Eulerian and Lagrangian frame depends on the displacement field $\fett{\Psi}(\fett{q},\eta)$, 
\be
 J^F(\fett{q},\eta)  \equiv \det \left[ \delta_{ij}+\Psi_{i,j} \right] 
  = 1 + \Psi_{i,i} + \frac 1 2 \left[ \Psi_{i,i} \Psi_{j,j} - \Psi_{i,j}\Psi_{j,i} \right] + \det \left[ \Psi_{i,j} \right] \, .
\ee
Note that this is an exact relation, i.e., valid for the exact displacement field $\fett{\Psi}$.
As stated above, the exact $\fett{\Psi}(\fett{q},\eta)$ is expanded in a series. 
As we shall see, the spatial parts
of the perturbations agree with the spatial parts in Eq.\,(\ref{pert1}) at each order, so we keep the 
previously introduced notation for them, but relabel the temporal parts to
$D(\eta)$, $E(\eta)$, etc.:
\begin{eqnarray} \label{pert2}
  \fett{\Psi}(\fett{q},\eta) = \epsilon D(\eta)\,  \fett{\Psi}^{(1)}(\fett{q}) + \epsilon^2 E(\eta)\,  \fett{\Psi}^{(2)}(\fett{q}) 
   &+& \epsilon^3 F(\eta)\,  \fett{\Psi}^{(3)}(\fett{q}) +\epsilon^4  G(\eta) \, \fett{\Psi}^{(4)}(\fett{q})   \nonumber \\ 
   &+& \epsilon^3 F_T(\eta)\,  \fett{T}^{(3)}(\fett{q})+\epsilon^4 G_T(\eta)\,\fett{T}^{(4)}(\fett{q})\, .
\end{eqnarray}
In the following we suppress the explicit temporal and spatial dependences. 
From Eq.\,(\ref{pert2}) we can approximate the Jacobian by:
\begin{eqnarray}
\label{Jacobian} J^F(\fett{q},\eta) &=& 1 + \epsilon D\,\mu_1^{(1)}  +\epsilon^2 D^2\,\mu_2^{(1)}  
  + \epsilon^2 E\,\mu_1^{(2)}  + \epsilon^3 F\,\mu_1^{(3)} + 2\epsilon^3 DE \,\mu_2^{(1,2)}
  \nonumber \\ &&\qquad+ \epsilon^3 D^3\,\mu_3^{(1)} +2\epsilon^4 DF \, \mu_2^{(1,3)}   
  - \epsilon^4 DF_T \Psi_{i,j}^{(1)}T_{j,i}^{(3)}  \nonumber \\ &&\qquad +\epsilon^4 G\,\mu_1^{(4)}
   +  \epsilon^4 E^2\,\mu_2^{(2)} 
 +\frac 1 2 \epsilon^4 D^2 E \, \varepsilon_{ikl} \varepsilon_{jmn} \Psi_{k,m}^{(1)}\Psi_{l,n}^{(1)}\Psi_{j,i}^{(2)} \,,
\end{eqnarray}
and the $a$th scalars $ \mu_{a}^{(n)} \equiv \mu_{a}^{(n)} (\fett{q})$ are defined by
\begin{align} \label{mu}
 \mu_{1}^{(n)} &\equiv {\Psi}_{i,i}^{(n)} \, ,\\
 \mu_2^{(n)}   &\equiv \frac 1 2 \left(  {\Psi}_{i,i}^{(n)} {\Psi}_{j,j}^{(n)}
      - {\Psi}_{i,j}^{(n)} {\Psi}_{j,i}^{(n)} \right) \,, \\
 \mu_3^{(n)}   &\equiv \det \left[ {\Psi}_{i,j}^{(n)}  \right] = 
   \frac 1 6 \,\varepsilon_{ikl} \, \varepsilon_{jmn} \, \Psi_{j,i}^{(n)} \Psi_{m,k}^{(n)} \Psi_{n,l}^{(n)} \,,
\intertext{and specifically}
 \mu_2^{(m,n)} &\equiv  
\frac 1 2 \left(  {\Psi}_{i,i}^{(m)} {\Psi}_{j,j}^{(n)} -  {\Psi}_{i,j}^{(m)} {\Psi}_{j,i}^{(n)} \right) \,.
\end{align}
Note that $\mu_2^{(m,n)} = \mu_2^{(n,m)}$ for any tensor ${\Psi}_{i,j}^{(n)}$ and
not only for longitudinal fields as pointed out incorrectly 
in \cite{Catelan:1994ze}, since it only consists 
of interchangable dummy indices.

The above scalars will also be used for the spatial parts of the $n$th-order perturbations in $\fett{p}$, since they are identical.

%%%%%%%%%%%%%%%%%%%%%%%%%%%%%%%%%%%%%%%%%%%%%%%%%%%%%%%%%%%%%%%%%%%
%%%%%%%%%%%%%%%%%%%%%%%%%%%%%%%%%%%%%%%%%%%%%%%%%%%%%%%%%%%%%%%%%%%
\subsection{The perturbation equations in Lagrangian
  form for an EdS background}\label{secPert}

As before, we start with the perturbation equations in the full-system approach. 
We first concentrate on solving the source equation, Eq.\,(\ref{LN3fin}),
and write down the perturbative irrotationality condition 
for the velocity field, Eq.\,(\ref{LN1fin}).
 Afterwards, in section \ref{secPert2}, we perform the 
same steps for the peculiar-system.

\subsubsection{Full-system approach}\label{secPert1}

Inserting the ansatz, Eq.\,(\ref{pert1}) and Eq.\,(\ref{iniRho}) into Eq.\,(\ref{LN3fin}), we 
obtain the following set of equations to be solved:
\begin{align}
  \label{pertEqFin1} &\epsilon^0 \left\{ 4\pi G \overline{\rho}_0 + 3 \ddot a a^2  \right\} = 0 \,, \\
  \label{pertEqFin2} &\epsilon^1 \left\{ \left[ a^2 \ddot{q}_1 +2 \ddot a a q_1 \right] \mu_1^{(1)}+ 4\pi G \overline{\rho}_0 \delta_0   \right\} = 0  \,, \\
   \label{pertEqFin3} &\epsilon^2 \left\{ \left[ a^2 \ddot{q}_2 + 2 \ddot a a q_2  \right] \mu_1^{(2)} 
   + \left[ 2 q_1 \ddot{q}_1 a+ \ddot a q_1^2 \right] \mu_2^{(1)}    \right\}  = 0 \,, \\
   \label{pertEqFin3.1}  &\epsilon^3 \left\{ \left[ a^2 \ddot{q}_3 + 2 \ddot a a q_3  \right] \mu_1^{(3)} 
  + 2 \left[ \ddot a q_1 q_2+ a q_1 \ddot{q}_2 + a \ddot{q}_1 q_2   \right] \mu_2^{(1,2)} + 3 \ddot{q}_1 q_1^2  \mu_3^{(1)} \right\} \hspace{0.39cm}= 0 \,, \\
 \label{pertEqFin4}  &\epsilon^4 \;\Big\{  \hskip-0.1cm\left[ a^2 \ddot{q}_4 + 2 \ddot a a q_4  \right] \mu_1^{(4)}
  +   \left[ \ddot a q_2^2 + 2 a \ddot{q}_2 q_2  \right] \mu_2^{(2)}  
 + \varepsilon_{ilm} \varepsilon_{jpq} \Big[ \ddot{q}_2 \frac{q_1^2}{2} 
     + \ddot{q}_1 q_1 q_2 \Big] \Psi_{p,l}^{(1)} \Psi_{q,m}^{(1)} \Psi_{j,i}^{(2)} \Big. \nonumber \\
 & \ws \vspace{0.1cm}+  \Big. 2 \left[ \ddot a q_1 q_3 +a \ddot{q}_1 q_3 + a q_1 \ddot{q}_3 \right] \mu_2^{(1,3)}
  -   \left[ \ddot a q_1 q_3^T + a \ddot{q}_1 q_3^T +a q_1 \ddot{q}_3^T  \right] \Psi_{i,j}^{(1)} T_{j,i}^{(3)} 
  \Big\}  = 0 \,.
\end{align}
Note that in the fourth-order part there is the occurrence of the transverse perturbation $\fett{T}^{(3)}$; thus, in order to
solve this equation we have to constrain it with the irrotationality condition. Inserting the ansatz 
into Eq.\,(\ref{LN1fin}),  the third- and fourth-order parts are:
\begin{eqnarray}
 \label{irrot1}  \epsilon^3  \left\{ \left[ a \dot{q}_3^T - \dot a q_3^T \right] \varepsilon_{ijk} T_{k,j}^{(3)}
   - \left[ q_1 \dot{q}_2 - \dot{q}_1 q_2  \right] \varepsilon_{ijk} \Psi_{l,j}^{(1)} \Psi_{l,k}^{(2)}  \right\} &=& 0 \,,  \\
 \label{irrot2} \epsilon^4  \left\{ \left[ a \dot{q}_4^T - \dot a q_4^T \right] \varepsilon_{ijk} T_{k,j}^{(4)}
   - \left[ q_1 \dot{q}_3 - \dot{q}_1 q_3  \right] \varepsilon_{ijk} \Psi_{l,j}^{(1)} \Psi_{l,k}^{(3)}  \right. \,\,\; &\ws& \nonumber \\
  \ws\hspace{4cm} - \left. \left[ q_1 \dot{q}_3^T - \dot{q}_1 q_3^T  \right] \varepsilon_{ijk} \Psi_{l,j}^{(1)} T_{l,k}^{(3)} \right\} &=& 0 \,.
\end{eqnarray}
As mentioned above, the irrotationality of the acceleration field follows from the irrotationality of the velocity field, 
as can be easily checked by time differentiating Eq.\,(\ref{irrot1}) and Eq.\,(\ref{irrot2}) and comparing it directly with the 
perturbation equation of Eq.\,(\ref{LN2fin}).

\subsubsection{Peculiar-system approach}\label{secPert2}

The peculiar equations, by construction, start
at the order ${\cal O}(\epsilon^1)$.
Inserting the ansatz, Eq.\,(\ref{pert2}), and the Jacobian, Eq.\,(\ref{Jacobian}), 
into the source equation Eq.\,(\ref{LN6fin}), delivers:
\begin{eqnarray} 
\label{PecpertEqFin1} &&\epsilon^1 \left\{ \left[ D'' - \alpha D \right] \mu_1^{(1)}  \right\} = 0  \,, \\
\label{PecpertEqFin2} &&\epsilon^2 \left\{ \left[ E''- \alpha E \right] \mu_1^{(2)} 
   - \left[ \alpha D^2 - 2D D'' \right] \mu_2^{(1)}    \right\}  = 0 \,, \\
\label{PecpertEqFin3} &&\epsilon^3 \left\{ \left[ F''- \alpha F \right] \mu_1^{(3)} 
  + 2  \left[ D''E+DE''- \alpha DE \right] \mu_2^{(1,2)} \right.  \nonumber \\
    &&\ws\hspace{7.57cm} + \left. \left[ 3 D'' D^2- \alpha D^3 \right]  \mu_3^{(1)} \right\}  = 0 \,, \\
\label{PecpertEqFin4} &&\epsilon^4 \;\Big\{  \hskip-0.1cm\left[ G''- \alpha G \right] \mu_1^{(4)}
  +   \left[ 2 E'' E - \alpha E^2 \right] \mu_2^{(2)}  
  \Big. \nonumber \\
 &&\ws\hspace{0.1cm}+ \varepsilon_{ilm} \varepsilon_{jpq} \Big[ \frac{D^2 E''}{2}+D''D E 
         - \alpha \frac{D^2 E}{2} \Big] \Psi_{p,l}^{(1)} \Psi_{q,m}^{(1)} \Psi_{j,i}^{(2)} \nonumber \\
 &&\ws\hspace{0.1cm}
   \Big.  +  2 \left[ D''F+DF'' +\alpha DF \right] \mu_2^{(1,3)} 
  - \left[ D''F_T+DF_T'' -\alpha DF_T \right] \Psi_{i,j}^{(1)} T_{j,i}^{(3)}   \Big\}  = 0 \,.
\end{eqnarray}
With the irrotationality condition of the peculiar-velocity, Eq.\,(\ref{LN4fin}), we obtain:
\begin{eqnarray}
 \label{irrot1pec} &&\epsilon^3  \left\{ F_T' \varepsilon_{ijk} T_{k,j}^{(3)}
   - \left[ DE' -D'E  \right] \varepsilon_{ijk} \Psi_{l,j}^{(1)} \Psi_{l,k}^{(2)}  \right\} = 0 \,,  \\
 \label{irrot2pec} &&\epsilon^4  \left\{ G_T' \varepsilon_{ijk} T_{k,j}^{(4)}
   - \left[ DF'-D'F\right] \varepsilon_{ijk} \Psi_{l,j}^{(1)} \Psi_{l,k}^{(3)} \right. \nonumber \\
   &&\ws\hspace{2.04cm}
   - \left. \left[ DF_T'- D'F_T  \right] \varepsilon_{ijk} \Psi_{l,j}^{(1)} T_{l,k}^{(3)} \right\} = 0 \,.
\end{eqnarray}
Again, one may obtain the irrotationality of the peculiar-acceleration by time differentiating Eqs.\,(\ref{irrot1pec}-\ref{irrot2pec}). 

%%%%%%%%%%%%%%%%%%%%%%%%%%%%%%%%%%%%%%%%%%%%%%%%%%%%%%%%%%%%%%%%%%%%
\subsection{Remark}\label{secComp}

Before we discuss the solutions of the aforementioned sets of equations we wish to say a few words on their usage in the current 
literature.

Eqs.\,(\ref{LN4fin}-\ref{LN6fin}) are also reported in \cite{Catelan:1994ze},
however the irrotationality condition for the peculiar-velocity 
(and hence for the peculiar-acceleration as well) is flawed. 
Specifically, once the tensor $T_{i,j}^{(n-1)}$ is nonzero, 
their expression for $\fett{T}^{(n)}$ is false. $T_{i,j}^{(n-1)}(\fett{q})\neq \fett{0}$  means
that the inhomogeneous deformation tensor $\Psi_{i,j}(\fett{q},\eta)$ consist also of an antisymmetric part,
i.e., in the most general case $\fett{T}^{(2)} \neq \fett{0}$ and thus $\fett{T}^{(3+k)}$
is not correct, with $k \in {\mathbb{N}}_0$.

The corrected irrotationality condition for the peculiar-velocity  in \cite{Catelan:1994ze} must read: 
\be \label{irrotCat}
 \varepsilon_{ijk} \left[ \left( 1+ {\Psi_{n,n}} \right) \delta_{lj}
 - \Psi_{l,j} +\Psi_{l,j}^c \right] {\Psi}_{k,l}' = 0 \,,  \hspace{1.5cm} (i = 1,2,3) \,\,.
\ee
Eg.\,(\ref{irrotCat}) is equivalent to Eq.\,(\ref{LN4fin}), as can be seen by decomposing 
an arbitrary tensor $\Psi_{i,j}$ into a symmetric and an antisymmetric part and evaluating both equations. 
Note that the subsequent perturbative calculation simplifies clearly with the usage of Eq.\,(\ref{LN4fin}).

Finally, we would like to stress again, that \emph{Lagrangian} and \emph{Eulerian}
transverse motions are \emph{not} the same, since both frames are connected by a non-linear
transformation (see Eq.\,\ref{mc}). In reference \cite{Lachieze:1993} the Eulerian irrotationality condition was 
not solved, and as a \emph{wrong} consequence they concluded that the Lagrangian 
transversality is zero at all orders, as long as the motion is irrotational in the Eulerian frame.
In order to avoid confusion with this issue in the following, we shall refer to `{irrotational}' only with respect to the Eulerian frame, whereas we reserve `{transverse}' to the Lagrangian frame.

%%%%%%%%%%%%%%%%%%%%%%%%%%%%%%%%%%%%%%%%%%%%%%%%%%%%%%%%%%%%%%%%%%%
\section{From general initial conditions to `Zel'dovich type' initial conditions}\label{secIC}

The general solution in Lagrangian perturbation theory consists of solving the above set of equations with (3+1) initial condition functions. However, 
for an irrotational fluid we have to specify only two initial potentials, one for the peculiar-velocity $\fett{u}(\fett{q},t_0) \!\equiv\! \boldsymbol{\nabla}_{\fett{q}}\,{\cal S}_1(\fett{q},t_0)$, and the other for the peculiar-acceleration $\fett{g}_{{\rm pec}}(\fett{q},t_0) \!\equiv \!\boldsymbol{\nabla}_{\fett{q}}\,{\cal S}_2(\fett{q},t_0)$ (where the latter is linked to the density contrast via the Poisson equation). 

In \cite{Buchert:1992ya} it has been shown that under the consideration 
of the most general ansatz for the integral curves of the full-velocity field $\fett{f}$, one may express the solutions in terms of the general initial conditions setting. 
For example, in the full-system approach the first-order integral curves $\fett{f}^{(1)}$ read in terms of the initial conditions:
\begin{align} \label{integral}
 \fett{f}^{(1)} &= a \fett{q} 
   +b_1(t) \,\fett{u}^D(\fett{q},t_0) \,t_0 +b_2(t) \, \fett{u}^R(\fett{q},t_0) \,t_0 
   +b_3(t)\, \fett{g}_{{\rm pec}}^D(\fett{q},t_0) \,t_0^2 \,,
\end{align}
where we have decomposed the peculiar-velocity in a curl-free part $\fett{u}^D$ and a divergence-free part $\fett{u}^R$, i.e., $\fett{u}\!=\! \fett{u}^D\!+\!\fett{u}^R$, and similar for $\fett{g}_{{\rm pec}}$, but  in this formulation the vortical part of $\fett{g}_{{\rm pec}}$ is zero, thus $\fett{g}_{{\rm pec}}\!=\! \fett{g}_{{\rm pec}}^D$. 
The explicit time coefficients $b_i(t)$ are not needed for this demonstration, but see Eq.\,(23a) in  \cite{Buchert:1992ya}. 

To avoid formally lengthy expressions for the solutions we impose a functional dependence on the two initial potentials ${\cal S}_1$ and ${\cal S}_2$.
The simplest classes of such functional constraints, that we call `Zel'dovich-type' initial conditions, only prescribe one initial potential \cite{Buchert:1993xz}, e.g. ${\cal S}_1= {\cal S}_2 \equiv {\cal S}$. 
In the following, we describe two different settings, which are consistent within the Zel'dovich approximation and are thus appropriate for studying the inhomogeneities of the 
large-scale structure: the \textit{slaved initial conditions} and the \textit{inertial initial conditions}.

\paragraph{Slaved initial conditions.} 

We require the following parallelity condition between the 
initial peculiar-velocity $\fett{u}(\fett{q},t_0)$ and the initial 
peculiar-acceleration $\fett{g}_{\rm pec }(\fett{q},t_0)$ (note that, in general, parallelity allows for an arbitrary scalar proportionality function):
\be \label{IC}
  \fett{u}(\fett{q},t_0) = \fett{g}_{\rm pec} (\fett{q},t_0)\, t_0 \;.
\ee
From this it follows that $\fett{u}^D(\fett{q},t_0)= \fett{g}_{\rm pec}^D (\fett{q},t_0)\, t_0$ and
$\fett{u}^R(\fett{q},t_0)= \fett{g}_{\rm pec}^R (\fett{q},t_0)\, t_0 =\fett{0}$. Thus, the integral 
curves~(\ref{integral}) reduce to the expression
\be
 \fett{f}_{Z1}^{(1)} = a \fett{q}+ \left[ b_1(t)+b_3(t) \right]\, \fett{u}^D(\fett{q},t_0)\,t_0 \,.
\ee
Since $\fett{g}_{\rm pec} (\fett{q},t_0)$ is non-vanishing, we have $\boldsymbol{\nabla}_{\fett{q}} \cdot \fett{g}_{\rm pec} (\fett{q},t_0)\propto \delta_0$ initially, where the initial density contrast figures in the exact expression $\delta = (1+\delta_0)/J^F-1$.

In \cite{Buchert:1993xz,Buchert:1993ud,Bildhauer:1991mj} it has been 
pointed out that the above restriction is appropriate for modelling the large-scale
structure: gravitational instability supports the tendency for density perturbations to
grow along potential flows---as long as one restricts the problem to irrotational flows. 
These so-called \emph{slaved} initial conditions are thus a dynamical outcome of the Eulerian linear perturbation theory,
where the growing mode supports the above parallelity condition. Consequently, this class of initial conditions is well-motivated, and under this initial assumption parallelity is then exactly preserved by the first-order Lagrangian solution. 
Physically, this setting implies initial density- and velocity perturbations, which is in accordance with the prediction of standard inflationary theories. This initial condition is also commonly used in $N$-body simulations.

With the constraint  Eq.\,(\ref{IC}) we only need to specify one potential initially, and
set $\fett{u}(\fett{q},t_0) := \boldsymbol{\nabla}_{\fett{q}} {\cal S}(\fett{q},t_0)$ and,
thus,
\be \label{gpec}
 \fett{g}_{\rm pec} (\fett{q},t_0) = \boldsymbol{\nabla}_{\fett{q}} {\cal S}(\fett{q},t_0) \,t_0 \,. 
\ee
According to the initial density-perturbation which is specified to be 
$\rho_0 = \overline{\rho}_0 \left( 1+ \epsilon \delta_0 \right)$ we have to relate
the initial potential ${\cal S}(\fett{q},t_0)$ to the potential $\phi^{(1)}(\fett{q})$ of the 
longitudinal perturbations: 
$\fett{\Psi}^{(1)}(\fett{q}) \equiv \boldsymbol{\nabla}_{\fett{q}} \phi^{(1)}(\fett{q})$.
We shall do so by using Poisson's equation for Eq.\,(\ref{gpec}). 
The slaved initial conditions shall be used in the following section~\ref{secSol1}.

\newpage

\paragraph{Inertial initial conditions.} Another class of initial conditions is to require that
\be \label{IC2}
  \fett{g}_{\rm pec}(\fett{q},t_0) = \fett{0} \,,
\ee
which can be considered as a `Zel'dovich-type' class as well. 
From this it follows that $\fett{u}^R(\fett{q},t_0)= \fett{0}$ and $\fett{g}_{\rm pec}^D (\fett{q},t_0)= \fett{0}$. Then, the integral 
curves~(\ref{integral}) become
\be
 \fett{f}_{Z2}^{(1)} = a \fett{q}+ b_1(t)\, \fett{u}^D(\fett{q},t_0)\,t_0 \,.
\ee
We call this class inertial initial conditions, 
since the initial acceleration is vanishing. 
In contrast to the slaved initial conditions this implies that 
there is no initially prescribed density perturbation, because Poisson's equation dictates a 
vanishing source term initially (see Eq.\,(\ref{PoissonGpec})). 
Instead we demand an initial homogeneous background density $\overline{\rho}(t_0)$, and the evolving density perturbations
are produced solely by initial velocity perturbations \cite{Buchert:1993xz}.  
As before we can then
set initially the peculiar-velocity to $\fett{u}(\fett{q},t_0) := \boldsymbol{\nabla}_{\fett{q}}\,{\cal S}(\fett{q},t_0)$.
In this case, mass conservation is given by the \emph{exact} setting $\delta = 1/J^F-1$.\footnote{With slaved initial conditions we may, however, assume an initial quasi-homogeneity, and hence approximate mass conservation by $\delta \simeq 1/J^F-1$, because of numerical smallness of the initial conditions. Alternatively, we may choose a different set of Lagrangian coordinates to render this form of the integral exact, but this will introduce formal complications. Zel'dovich's original model implicitly supposed slaved initial conditions, while the initial density contrast was approximated by zero in the density integral.} We shall use inertial initial conditions to solve the equations in the peculiar-system approach (see section~\ref{secSol2}).

Although inertial initial conditions are of the `Zel'dovich-type', it is difficult to 
justify their physical motivation. The main reason is because it is nowadays believed that the CMB fluctuations are of adiabatic nature, and the (non-relativistic) counter-part of adiabatic initial perturbations would rather correspond to slaved initial conditions. However, as we shall see in the upcoming sections, only decaying modes change, whereas the fastest growing modes coincide with the use of both initial conditions.

In an expanding universe, the physical significance of the decaying modes is usually negligible, because they are dominated by the fastest growing mode. However, in the highly non-linear regime, decaying modes can affect the clustering via their coupling to growing modes \cite{Buchert:1993ud}. (Also, since the equations and solutions are time-reversable, their application backward in time exchanges the role of the importance of growing and decaying modes.)
Restricting the general initial conditions setting is mainly motivated by simplification. There is a price to pay for these restrictions when one is interested in the structural details 
at highly non-linear stages at small scales, but this regime lies in the realm of $N$-body simulations. 

%%%%%%%%%%%%%%%%%%%%%%%%%%%%%%%%%%%%%%%%%%%%%%%%%%%%%%%%%%%%%%%%%%%
\section{Solutions of the perturbation equations 
  for an EdS background: \\full-system approach}\label{secSol1}

In this section we solve the perturbation equations with the use of the slaved initial conditions, see Eq.\,(\ref{IC}).

\setcounter{subsection}{-1}

\subsection{The zero-order solution}

To find the solution for the homogeneous and isotropic background, Eq.\,(\ref{pertEqFin1}), 
one has to solve the Friedmann equation for the cosmic scale factor $a(t)$. 
In the case of an EdS
universe we have:
\begin{eqnarray}
  a(t) = \left( \frac{t}{t_0} \right)^{2/3} \;.
\end{eqnarray}
Using the Friedmann equation again we can write the prefactor $4 \pi G \overline{\rho}_0$ 
in Poisson's equation as $2/(3t_0^2)$.

\subsection{The first-order solution}\label{secFirst}

Using the above result and the longitudinal ansatz  $\fett{\Psi}^{(1)}(\fett{q}) \equiv \fett{\nabla_q} \phi^{(1)}(\fett{q})$, 
Eq.\,(\ref{pertEqFin2}) reads:
\be \label{FOS}
 \left[ \ddot{q}_1 +2  \frac{\ddot a}{a} q_1 \right] \Delta_{\fett{q}} \phi^{(1)} =  - \frac{2}{3 t_0^2 a^2} \delta_0   \,. 
\ee
Comparing with Poisson's equation for the initial potential
${\cal S}(\fett{q},t_0)$, namely
\be \label{PoissonGpec}
 \Delta_{\fett{q}} {\cal S} (\fett{q},t_0) / t_0 = -\frac{2}{3t_0^2} \delta_0  \equiv \fett{\nabla_q} \cdot \fett{g}_{ \rm pec} (\fett{q},t_0) \,,
\ee
we can read off the connection between the spatial dependence and the initial conditions. The solution is then
\begin{eqnarray}
  \left[ \ddot{q}_1 +2  \frac{\ddot a}{a} q_1 \right] = \frac{1}{a^2 t_0^2}  \quad;\quad
 \Delta_{\fett{q}} \phi^{(1)} (\fett{q}) \equiv \mu_1^{(1)} = \Delta_{\fett{q}} {\cal S} (\fett{q},t_0)\,t_0 \;\;.
\end{eqnarray}
The solution of the time-evolution under the restriction Eq.\,(\ref{IC}) can be found by the formal requirement $q_1(t_0) \!=\! 0$, and the condition that
the coefficient functions of the peculiar-velocity and peculiar-acceleration are equal to 1 at $t=t_0$ \cite{Buchert:1993ud}: $\dot{q}_1(t_0) = 1/t_0$. The temporal solution is then
\be
  q_1 = \frac 3 2 \left[ a^2 -a \right]  \,.
\ee

\subsection{The second-order solution}

After separating the spatial part from the temporal part of Eq.\,(\ref{pertEqFin3}) we have:
\begin{eqnarray}
 \left[ a^2 \ddot{q}_2 + 2 \ddot a a q_2  \right]   
    = - \left[ 2 q_1 \ddot{q}_1 a+ \ddot a q_1^2 \right]
     \quad;\quad  \mu_1^{(2)} = \mu_2^{(1)}   \;\;.
\end{eqnarray}
This linear ordinary differential equation can be solved and is found to be
\be
  q_2 = \left( \frac 3 2 \right)^2 \left[  -\frac{3}{7} a^3 + \frac 6 5 a^2 - a + \frac{8}{35} a^{-1/2} \right] \,,
\ee
with the restrictions $q_2(t_0)\! =\!0$ and $\dot{q}_2(t_0)\! =\!0$. The same restrictions are used for the higher-order solutions as well.

\subsection{The third-order solution}\label{3rdFull}

Eq.\,(\ref{pertEqFin3.1}) consists of two separate spatial parts with its temporal parts, the first of 
which describes a coupling between first- and 
second-order perturbations, the second is dependent on cubic first-order perturbations. 
As suggested  in \cite{Buchert:1993xz} the third-order 
displacement is split into two longitudinal parts, arising from these two specific dependencies and 
then solved piecewise.
Thus, the solution of Eq.\,(\ref{pertEqFin3.1}) consists in solving the two sets of equations, which we
label with $3a$ and $3b$. Additionally, we have the occurrence of a transverse solution at the third order 
from Eq.\,(\ref{irrot1}), 
which we label with $3c$ and separate it into spatial and temporal parts as well. The three sets of equations then read:
\begin{align}  
   \left[ \ddot{q}_{3a} + 2 \frac{\ddot a}{a} q_{3a}  \right] = - 3 \ddot{q}_1 \frac{q_1^2}{a^2} &\quad;\quad 
\mu_1^{(3a)} = \mu_3^{(1)}  \,,  \nonumber \\
 \left[ \ddot{q}_{3b} + 2 \frac{\ddot a}{a} q_{3b} \right] = 
    - 2 \left[ \frac{\ddot a}{a^2} q_1 q_2+ q_1 \frac{\ddot{q}_2}{a} + \frac{\ddot{q}_1}{a} q_2 \right]
 &\quad;\quad    \mu_1^{(3b)} = \mu_2^{(1,2)}  \,, \\
   \left[ \dot{q}_{3c}^T - \frac{\dot a}{a} q_{3c}^T \right] 
    =  \left[ \frac{q_1}{a} \dot{q}_2 - \frac{\dot{q}_1}{a} q_2  \right] 
  &\quad;\quad  
 \varepsilon_{ijk} T_{k,j}^{(3c)} =  \varepsilon_{ijk} \Psi_{l,j}^{(1)} \Psi_{l,k}^{(2)}
  \;. \nonumber
\end{align}
As mentioned earlier the purely longitudinal parts are $\mu_1^{(n)} \equiv \Delta_{\fett{q}} \phi^{(n)}$. The
purely transverse part can be written in terms of a vector potential, i.e., $T_k^{(3c)} \equiv \varepsilon_{klm} \partial_{q_l} A_m^{(3c)}$,
where one degree of freedom may be cancelled by employing the Coulomb gauge.
Using the same restrictions as before, we obtain for the temporal parts of the three above equations:
\begin{eqnarray}
 \begin{array}{ll} 
 q_{3a} &= \left(\frac 3 2 \right)^{3} \left[ -\frac 1  3 a^4 + \frac 9 7 a^3 - \frac 9 5 a^2 + a -\frac{16}{105}a^{-1/2}  \right]  \,, \\ \textcolor{white}{a}  \\ 
  q_{3b} &= \left(\frac 3 2 \right)^{3} \left[ \frac {10}{21} a^4 - \frac{66}{35} a^3 + \frac {14}{5} a^2 - 2 a + \frac{16}{35} a^{1/2} +\frac{16}{105}a^{-1/2}  \right] \,,  \\   \textcolor{white}{a} \\
  q_{3c}^T &= \left(\frac 3 2 \right)^{3} \left[ -\frac {1}{7} a^4 + \frac{3}{7} a^3 + \frac {1}{5} a^2 - a + \frac{8}{7} a^{1/2} -\frac{8}{35}a^{-1/2}  \right] \,.
  \end{array}   
\end{eqnarray}

\subsection{The fourth-order solution}

The fourth-order part of the source equation, Eq.\,(\ref{pertEqFin4}), consists of five separate spatial 
parts with their respective temporal parts. Analogous to the last section we label them with $4a - 4e$.
The fourth-order part of the irrotationality condition, Eq.\,(\ref{irrot2}), consists  of two \textit{types}
of spatial parts, the first type arises from longitudinal perturbations only, i.e., 
$\varepsilon_{ijk} \Psi_{l,j}^{(1)} \Psi_{l,k}^{(3)}$, whereas the second type
results from a mixing term of longitudinal-transverse perturbations, i.e.,
 $\varepsilon_{ijk} \Psi_{l,j}^{(1)} T_{l,k}^{(3)}$. Due to the fact that the third-order longitudinal 
perturbations include the solutions $3a$ and $3b$, we obtain two terms contributing to the first type, which 
we label with $4f$ and $4g$. The second type occurs only once which we label with $4h$.
The eight sets of equations to be solved then read:
\begin{eqnarray*} 
\left[ \ddot{q}_{4a} + 2 \frac{\ddot a}{a} q_{4a}  \right] 
  = - \left[ \frac{\ddot a}{a^2} q_2^2 + 2 \frac{\ddot{q}_2}{a} q_2  \right]
&&\quad;\quad \mu_1^{(4a)} = \mu_2^{(2)} \,,              \\ 
   \left[ \ddot{q}_{4b} + 2 \frac{\ddot a}{a} q_{4b}  \right] = 
    -  \Big[ \frac{\ddot{q}_2}{2a^2}{q_1^2} 
     + \frac{\ddot{q}_1}{a^2} q_1 q_2 \Big]
  &&\quad;\quad \mu_1^{(4b)} = \varepsilon_{ilm} \varepsilon_{jpq}\Psi_{p,l}^{(1)} \Psi_{q,m}^{(1)} \Psi_{j,i}^{(2)}  \, ,     \\
    \left[  \ddot{q}_{4c} + 2 \frac{\ddot a}{a} q_{4c} \right] 
    = -2 \left[ \frac{\ddot a}{a^2} q_1 q_{3a} + \frac{\ddot{q}_1}{a} q_{3a} +  \frac{q_1}{a} \ddot{q}_{3a} \right] 
  &&\quad;\quad \mu_1^{(4c)} =  \mu_2^{(1,3a)}  \, ,    \\
    \left[ \ddot{q}_{4d}  + 2 \frac{\ddot a}{a} q_{4d} \right] 
    = -2 \left[ \frac{\ddot a}{a^2} q_1 q_{3b} + \frac{\ddot{q}_1}{a} q_{3b} +  \frac{q_1}{a} \ddot{q}_{3b} \right] 
  &&\quad;\quad \mu_1^{(4d)} =  \mu_2^{(1,3b)}  \, ,    \\
    \left[ \ddot{q}_{4e} + 2 \frac{\ddot a}{a} q_{4e} \right] 
    = \left[ \frac{\ddot a}{a^2} q_1 q_3^T + \frac{\ddot{q}_1}{a} q_3^T +\frac{q_1}{a} \ddot{q}_3^T  \right] 
&&\quad;\quad\mu_1^{(4e)} = \Psi_{i,j}^{(1)} T_{j,i}^{(3c)}   \, ,  
\end{eqnarray*}
and
\begin{eqnarray}
    \left[  \dot{q}_{4f}^T - \frac{\dot a}{a} q_{4f}^T \right] 
    =  \left[ \frac{q_1}{a} \dot{q}_{3a} - \frac{\dot{q}_1}{a} q_{3a}  \right] 
  &&\quad;\quad \varepsilon_{ijk} T_{k,j}^{(4f)} =  \varepsilon_{ijk} \Psi_{l,j}^{(1)} \Psi_{l,k}^{(3a)}  \,,   \nonumber\\
    \left[ \dot{q}_{4g}^T - \frac{\dot a}{a} q_{4g}^T \right] 
    =  \left[ \frac{q_1}{a} \dot{q}_{3b} - \frac{\dot{q}_1}{a} q_{3b}  \right]
  &&\quad;\quad  \varepsilon_{ijk} T_{k,j}^{(4g)} =  \varepsilon_{ijk} \Psi_{l,j}^{(1)} \Psi_{l,k}^{(3b)}  \,,    \\
    \left[ \dot{q}_{4h}^T - \frac{\dot a}{a} q_{4h}^T \right] 
    =  \left[ \frac{q_1}{a} \dot{q}_{3c}^T - \frac{\dot{q}_1}{a} q_{3c}^T  \right] 
&&\quad;\quad \varepsilon_{ijk} T_{k,j}^{(4h)} =  \varepsilon_{ijk} \Psi_{l,j}^{(1)} T_{l,k}^{(3c)}  \, .  \nonumber 
\end{eqnarray} 

\noindent Using the same restrictions as before, we obtain for the temporal parts of the eight sets of equations:
\begin{eqnarray}
 \begin{array}{ll} q_{4a} &= \left(\frac 3 2 \right)^{4} \left[ -\frac{51}{539} a^5 + \frac 4 7 a^4 
     - \frac{258}{175} a^3 + \frac{12}{5} a^2 - \frac{48}{49}a^{3/2} \right.  \\ 
   &\left.\ws \hspace{4cm} -a+\frac{96}{175} \sqrt{a}+\frac{16}{385} a^{-1/2} -\frac{16}{1225} a^{-2}   \right]  \,, \\ \textcolor{white}{a}  \\ 
  q_{4b} &= \left(\frac 3 2 \right)^{4} \left[\frac{13}{154} a^5- \frac{46}{105}a^4+\frac{33}{35} a^3 -\frac{6}{5}a^2 
     + \frac{12}{35}a^{3/2}+\frac 1 2 a - \frac{8}{35}\sqrt{a} -\frac{4}{1155}a^{-1/2} \right] \,,  \\   \textcolor{white}{a} \\
  q_{4c} &= \left(\frac 3 2 \right)^{4} \left[ \frac{14}{33} a^5-\frac{44}{21}a^4+\frac{144}{35}a^3-4a^2+2a -\frac{32}{105}\sqrt{a} -\frac{32}{231}a^{-1/2} \right] \,, \\   \textcolor{white}{a} \\
  q_{4d} &= \left(\frac 3 2 \right)^{4} \left[ -\frac{20}{33}a^5+\frac{64}{21}a^4-\frac{216}{35}a^3+\frac{32}{5}a^2 -4a+\frac{128}{105} \sqrt{a}+\frac{128}{1155}a^{-1/2} \right] \,, \\   \textcolor{white}{a} \\
  q_{4e} &= \left(\frac 3 2 \right)^{4} \left[ -\frac{1}{11}a^5+\frac{8}{21}a^4-\frac{18}{35}a^3+a-\frac{32}{35}\sqrt{a}+\frac{32}{231}a^{-1/2}  \right] \,, \\   \textcolor{white}{a} \\
 q_{4f}^T &= \left(\frac 3 2 \right)^{4} \left[ -\frac{1}{6}a^5+\frac{16}{21}a^4-\frac 9 7 a^3+\frac 4 5 a^2 + \frac 1 2 a - \frac{16}{21} \sqrt{a} +\frac{16}{105}a^{-1/2}  \right] \,, \\   \textcolor{white}{a} \\
  q_{4g}^T &= \left(\frac 3 2 \right)^{4} \left[ \frac{5}{21}a^5-\frac{116}{105}a^4+\frac{66}{35}a^3-\frac 4 5 a^2 -\frac{48}{35}a^{3/2}+a+\frac{32}{105}\sqrt{a}-\frac{16}{105}a^{-1/2} \right] \,, \\   \textcolor{white}{a} \\
  q_{4h}^T &= \left(\frac 3 2 \right)^{4} \left[ -\frac{1}{14}a^5+\frac{2}{7}a^4-\frac{3}{7}a^3+\frac 6 5 a^2 -\frac{24}{7}a^{3/2}+\frac 9 2 a-\frac{16}{7}\sqrt{a}+\frac{8}{35}a^{-1/2}  \right] \,.
\end{array} 
 \,
\end{eqnarray}

\section{Solutions of the perturbation equations 
  for an EdS background: \\peculiar-system approach}\label{secSol2}

As promised above, we solve the perturbation equations in this section with inertial initial
conditions, see Eq.\,(\ref{IC2}).

\subsection{The first-order solution}

The first-order part of the peculiar-source equation, Eq.\,(\ref{PecpertEqFin1}), is
\be \label{FO}
 D''(\eta)  \,\Delta_{\fett{q}} \phi^{(1)} = \alpha(\eta) D(\eta) \,\mu_1^{(1)} \,,
\ee
with $\alpha = 6/\eta^2$. 
We fix the constants of the temporal solution $D(\eta)$ such that the normalisation coincides with the one obtained by linear SPT. 
Formally this can be achieved by the requirement that $D(\eta_0) =0$, and that the coefficient function of the peculiar-velocity is initially equal to $-5$: $\fett{u}(\fett{q},\eta_0) = 1/a(\eta_0) D'(\eta_0) \fett{\nabla_q} \phi^{(1)}= -\frac{5}{\eta_0} \fett{\nabla_q} \phi^{(1)}$.
 The full solution is then
\begin{eqnarray}  \label{linGrowth}
  D =  a- a^{-3/2}  \quad;\quad  \Delta_{\fett{q}} \phi^{(1)} (\fett{q}) \equiv \mu_1^{(1)} = - \frac 1 5 \Delta_{\fett{q}} {\cal S} (\fett{q},\eta_0)\,\eta_0 \,,
\end{eqnarray}
with the initial condition $\fett{u}(\fett{q},\eta_0) = \fett{\nabla_q}{\cal S}(\fett{q},\eta_0)$.

\subsection{The second-order solution}

Using Eq.\,(\ref{FO}) we can rewrite the second-order solution of Eq.\,(\ref{PecpertEqFin2}) as 
\begin{eqnarray} \label{2ndSol}
   \mu_1^{(2)} = \mu_2^{(1)}  \quad;\quad
   E''- \alpha E = - \alpha D^2 \;\;.
\end{eqnarray}
Then we obtain
\be \label{2ndTime}
  E = -\frac 3 7 a^2 +\frac 5 4  a - 2 a^{-1/2}+\frac{10}{7}  a^{-3/2} -\frac 1 4 a^{-3} \,,
\ee
with the restrictions $E(\eta_0) \!=\! 0$ and $E'(\eta_0)\! =\! 0$. The same restrictions are used for the higher-order solutions as well.

\subsection{The  third-order solution}

Analogous to section Eq.\,(\ref{3rdFull}) the third-order solution consists of two 
longitudinal ($a$ and $b$) and one transverse solution ($c$). Using the lower-order results we can write
the solutions of Eq.\,(\ref{PecpertEqFin3}) and Eq.\,(\ref{irrot1pec}) as follows:
\begin{eqnarray} \label{3rdSol}
  {F}_a'' -\alpha F_a = -2\alpha D^3 &&\quad;\quad \mu_1^{(3a)} = \mu_3^{(1)} \,,
              \nonumber \\ 
   {F}_b'' - \alpha F_b = -2\alpha D \left[ E- D^2 \right] &&\quad;\quad \mu_1^{(3b)} = \mu_2^{(1,2)} \,,
             \\
   {F}_c'' = - \alpha D^3 &&\quad;\quad \varepsilon_{ijk}\, T_{k,j}^{(3c)}  =  \varepsilon_{ijk}\, \Psi_{l,j}^{(1)} \Psi_{l,k}^{(2)} \;. \nonumber 
\end{eqnarray}
For the three time-evolution functions we obtain under the same restrictions as above:
\be
 \begin{array}{ll} \label{3rdTime}
  F_a &= - \frac 1 3 a^3 +\frac{75}{11} a - 9 a^{1/2} +\frac{25}{3} a^{-3/2}
   -6 a^{-2} +\frac{2}{11} a^{-9/2} \,, \\
  F_b &= \sign\frac{10}{21} a^3 - \frac{15}{14} a^2 -\frac{25}{11} a +\frac{30}{7}a^{1/2}
  +\frac{5}{14}a^{-1/2}-\frac{100}{21} a^{-3/2}   \\  
    &\qquad+\frac 5 2 a^{-2} +\frac 5 7 a^{-3} -\frac{5}{22} a^{-9/2}  \,,  \\
  F_c^T &= -\frac{1}{7} a^3 +9 a^{1/2}+\frac{375}{28}a^{-1/2} -\frac 3 2 a^{-2} +\frac{1}{12} a^{-9/2}  -\frac{125}{6} \,. 
 \end{array}
\ee

\subsection{The  fourth-order solution}

The solutions of the source equation, Eq.\,(\ref{PecpertEqFin4}), and of the irrotationality condition, Eq.\,(\ref{irrot2pec}) can be written as follows:
\begin{eqnarray} \label{4thSol}
   {G}_a'' -\alpha G_a = \alpha \left[ 2D^2E-E^2 \right] &&\quad;\quad \mu_{1}^{(4a)} = \mu_2^{(2)} \,, \textcolor{white}{a} \nonumber \\ 
   {G}_b'' -\alpha G_b = \alpha \left[ \frac{D^4}{2} -D^2E \right] &&\quad;\quad  \mu_{1}^{(4b)} = \varepsilon_{ikl} \varepsilon_{jmn} \Psi_{k,m}^{(1)} \Psi_{l,n}^{(1)} \Psi_{j,i}^{(2)} \,,  \textcolor{white}{a}  \nonumber \\
   {G}_c'' -\alpha G_c = -2 \alpha \left[ DF_a -2D^4 \right] &&\quad;\quad \mu_{1}^{(4c)} = \mu_2^{(1,3a)}  \,, \textcolor{white}{a}  \nonumber \\
   {G}_d'' -\alpha G_d = -2 \alpha \left[ DF_b -2D^2E+2D^4 \right] &&\quad;\quad \mu_{1}^{(4d)} = \mu_2^{(1,3b)} \,,   \textcolor{white}{a}   \\ 
   {G}_e'' -\alpha G_e = -\alpha D^4 &&\quad;\quad \mu_{1}^{(4e)} = \Psi_{i,j}^{(1)}T_{j,i}^{(3c)}  \textcolor{white}{a} \,,  \nonumber \\
   {G_f^T}'' = -2 \alpha D^4 &&\quad;\quad \varepsilon_{ijk}\, T_{k,j}^{(4f)} = \varepsilon_{ijk}\,\Psi_{l,j}^{(1)} \Psi_{l,k}^{(3a)} \,,  \textcolor{white}{a}  \nonumber \\
   {G_g^T}''  = -2 \alpha D^2\left[ E-D^2 \right] &&\quad;\quad \varepsilon_{ijk}\, T_{k,j}^{(4g)} = \varepsilon_{ijk}\, \Psi_{l,j}^{(1)} \Psi_{l,k}^{(3b)} \,,   \textcolor{white}{a} \nonumber \\
   {G_h^T}'' = -\alpha D \left[ D^3 + F_c \right] &&\quad;\quad \varepsilon_{ijk}\, T_{k,j}^{(4h)} =  \varepsilon_{ijk}\, \Psi_{l,j}^{(1)} T_{l,k}^{(3c)} \;. \nonumber 
\end{eqnarray}

\noindent We then obtain for the temporal parts of above sets of equations:
\begin{eqnarray}
 \begin{array}{ll} 
 G_a &= -\frac{51}{539}a^4 +\frac{25}{42} a^3-\frac{75}{112}a^2 - 4 a^{3/2}+\frac{2125}{231}a-\frac{300}{49}a^{1/2}+\frac{25}{7}a^{-1/2} -\frac{51}{14}a^{-1} \\ 
 &\qquad-\frac{250}{231}a^{-3/2}+\frac{25}{8}a^{-2} -\frac{25}{49}a^{-3}-\frac{2}{3}a^{-7/2}+\frac{25}{77}a^{-9/2}-\frac{3}{112}a^{-6}  \,, \\ \textcolor{white}{a}  \\
 G_b &=   \sign\frac{13}{154} a^4-\frac{5}{24}a^3 - \frac 6 7 a^{3/2}+\frac{2375}{924}a -\frac{45}{28}a^{1/2}+\frac{27}{56}a^{-1}-\frac{125}{66}a^{-3/2} \\ 
  &\qquad +\frac{45}{28}a^{-2} -\frac{1}{12}a^{-7/2}-\frac{10}{77}a^{-9/2}+\frac{1}{28}a^{-6}  \,, \\ 
 \textcolor{white}{a}  \\ 
 G_c &=  \sign\frac{14}{33} a^4  -\frac{450}{77}a^2+ \frac{4}{3}a^{3/2}+\frac{125}{14}a+\frac{100}{33}a^{-1/2} -27 a^{-1}+\frac{1500}{77}a^{-3/2} \\
&\qquad +\frac{25}{6}a^{-3} -\frac{52}{11}a^{-7/2}+\frac{16}{77}a^{-6}  \,, \\ \textcolor{white}{a}  \\ 
 G_d &=  -\frac{20}{33} a^4 +\frac{25}{21}a^3+ \frac{150}{77}a^2+ \frac{80}{21}a^{3/2}-\frac{4000}{231}a+\frac{75}{7}a^{1/2}-\frac{1150}{231}a^{-1/2} +\frac{75}{7}a^{-1} \\
&\qquad+\frac{250}{231}a^{-3/2}-\frac{50}{7}a^{-2}-\frac{50}{21}a^{-3} +\frac{85}{33}a^{-7/2}+\frac{50}{77}a^{-9/2}-\frac{20}{77}a^{-6}  \,, \\ \textcolor{white}{a}  \\ 
 G_e &=  -\frac{1}{11} a^4 + 4 a^{3/2} -\frac{125}{21}a +9 a^{-1}-\frac{250}{33}a^{-3/2} +\frac{2}{3}a^{-7/2}-\frac{1}{21}a^{-6}  \,, \\  \textcolor{white}{a}  \\ 
 \label{4thTime} G_f^T &=  -\frac{1}{6} a^4 +  4 a^{3/2} +\frac{2500}{33}a^{-1/2}-36 a^{-1}+\frac{8}{7}a^{-7/2}-\frac{1}{11}a^{-6} -\frac{625}{14}   \,, \\ \textcolor{white}{a}  \\ 
 G_g^T &=  \sign\frac{5}{21} a^4 -\frac{5}{14}a^3-\frac{20}{7}a^{3/2} +\frac{45}{7}a^{1/2}-\frac{8125}{231}a^{-1/2} +\frac{225}{14}a^{-1}+\frac{45}{28}a^{-2} \\
&\qquad-\frac{5}{7}a^{-7/2}-\frac{5}{21}a^{-9/2}+\frac{5}{44}a^{-6}+\frac{625}{42}   \,, \\ \textcolor{white}{a}  \\ 
 G_h^T &=  -\frac{1}{14} a^4 -\frac{18}{7}a^{3/2} +\frac{125}{6}a -\frac{1125}{28}a^{1/2} +\frac{27}{2}a^{-1}-\frac{125}{6}a^{-3/2} \\
 &\qquad+\frac{375}{56}a^{-2}+\frac{29}{84}a^{-7/2}-\frac{1}{24}a^{-6} +\frac{625}{28} \,,  \\ 
\end{array}  \,
\end{eqnarray}
respectively.

The above solutions will be used in the following section. For our purpose, i.e., setting up the relation to SPT, only the fastest growing modes will be relevant.

%%%%%%%%%%%%%%%%%%%%%%%%%%%%%%%%%%%%%%%%%%%%%%%%%%%%%%%%%%%%%%%%%%%
\section{Practical realisation and exact relationship between LPT and SPT}\label{sec:realisation}

In this section we explicitly link the perturbative series between LPT and SPT. 
Note that we still restrict to an EdS universe and to the Zel'dovich-type 
subclasses of the general Lagrangian solutions. As mentioned above, these subclasses
consist of the slaved and the inertial initial conditions.

Commonly, the dynamical quantities of the irrotational solutions 
in SPT are the density contrast and the peculiar-velocity divergence, 
so it is convenient to use in the following the peculiar-system approach only. 
In the latter we 
call $\fett{\Psi}$ the (gravitational induced) displacement field.

\subsection{The LPT series in Fourier space}

We seek a series solution of the form
\be \label{FTseries}
  \tilde{\fett{\Psi}}(\fett{k},t) = \sum_{n=1}^{\infty} \left[ \tilde{\fett{\Psi}}^{(n)}(\fett{k},t)
    + \tilde{\fett{T}}^{(n)}(\fett{k},t) \right] \,,
\ee
where a tilde denotes Fourier quantities, and we explicitly neglect the perturbation parameter $\epsilon$. Eq.\,(\ref{FTseries}) is the complementary series of Eq.\,(\ref{pert2}) in Fourier space for the displacement field.
Then, using the solutions of section \ref{secSol2}, 
it is possible to write the Fourier transform of the spatial $n$th-order
perturbations $\tilde{\fett{\Psi}}^{(n)}(\fett{k})$ and $\tilde{\fett{T}}^{(n)}(\fett{k})$ 
times the fastest temporal parts $\propto D^n(t)$ as\footnote{$D(t)$ is the fastest growing mode solution of the linear growth function in the peculiar-system approach, see Eq.\,(\ref{linGrowth}). For an EdS universe it is $D(t)=a(t)$.}
\begin{eqnarray} \label{finish4thText}
  {\tilde{\fett{\Psi}}}^{(n)} (\fett{k},t) &=& -\ii D^n(t) \int \frac{\dd^3 p_1 \cdots \dd^3 p_n}{(2\pi)^{3n}} 
  \, (2\pi)^3 \delta_D^{(3)}(\fett{p}_{1\cdots n} -\fett{k} ) \, \nonumber \\
 &&\ws \hspace{2cm} \times \fett{S}^{(n)} (\fett{p}_1, \ldots, \fett{p}_n) 
  \, \tilde{\delta}^{(1)} (\fett{p}_1) \cdots \tilde{\delta}^{(1)} (\fett{p}_n) \,,
\end{eqnarray}
and
\begin{eqnarray} \label{finish4thTranseText}
  {\tilde{\fett{T}}}^{(n)} (\fett{k},t) &=& -\ii D^n(t) \int \frac{\dd^3 p_1 \cdots \dd^3 p_n}{(2\pi)^{3n}} 
  \, (2\pi)^3 \delta_D^{(3)}(\fett{p}_{1\cdots n} -\fett{k} ) \, \nonumber \\
 &&\ws \hspace{2cm} \times \fett{S}_T^{(n)} (\fett{p}_1, \ldots, \fett{p}_n) 
  \, \tilde{\delta}^{(1)} (\fett{p}_1) \cdots \tilde{\delta}^{(1)} (\fett{p}_n) \,,
\end{eqnarray}
where we have employed the shorthand notation $\fett{p}_{1\cdots n}= \fett{p}_1 + \fett{p}_2 + \cdots + \fett{p}_n$, $\fett{S}^{(n)}$ and $\fett{S}_T^{(n)}$ are symmetrised vectors  arising from longitudinal and transverse perturbations, respectively (see appendix \ref{app:Res}).
$\tilde{\delta}^{(1)}(\fett{p})$ is the Fourier transform of the linear density contrast evaluated at $t_0$.  We derive the above expressions in detail in appendix \ref{app:Prep}.

It is important to be conscious of the origin of the linear density contrasts on the RHS in Eqs.\,(\ref{finish4thText}-\ref{finish4thTranseText}), since they include a specific limit. To see this we revisit the mass conservation:
\be \label{deltajf}
  \delta(\fett{x},t) = \frac{1}{J^F(\fett{q},t)} -1 \,, \quad {\rm with} \quad \fett{x}(\fett{q},t) = \fett{q} + \fett{\Psi}(\fett{q},t) \,.
\ee
Linearisation of this (for {\it any} displacement exact) Lagrangian expression implies for the spatial part
\be \label{ZAlimit}
   \delta^{(1)}(\fett{x}) \approx - \Delta_{\fett{q}} \phi^{(1)}(\fett{q}) \,,
\ee
with $\fett{\Psi}^{(1)}(\fett{q}) \equiv \fett{\nabla_q} \phi^{(1)}(\fett{q})$. In the ZA, the displacement field is very small, so it is appropriate to use the limit $\fett{x} \approx \fett{q}$, however neglecting the inherent non-linearity of the Lagrangian approach in this step. Thus, the quantities $\fett{x}$ and $\fett{q}$ in Eq.\,(\ref{ZAlimit}) are interchangeable, and this reads in Fourier space
\be  \label{ZAft}
 \tilde{\delta}^{(1)}(\fett{p})  \approx p^2 \tilde{\phi}^{(1)} (\fett{p})  \,,
\ee
at the linear order.  
The $n$ linear density contrasts in ${\tilde{\fett{\Psi}}}^{(n)}$ and in ${\tilde{\fett{T}}}^{(n)}$  result from $n$ times the usage of Eq.\,(\ref{ZAft}). Stated in another way, ${\tilde{\fett{\Psi}}}^{(n)}$ and ${\tilde{\fett{T}}}^{(n)}$ are affected by this approximation at any order, but strictly speaking, only because the above linear approximation is used $n$ times.
We call this setting the \emph{initial position limit} (IPL).\footnote{Given an arbitrary tensor function $T(\fett{x},t)$ with the trajectory $\fett{F} \equiv \fett{x}(t) = \fett{q}+\Psi(\fett{q},t)$. Then the IPL is $T_{\rm IPL}(\fett{x},t) := \lim_{\fett{\fett{\Psi} \rightarrow \fett{0}}}T(\fett{x},t) = T(\fett{q},t)$, which is a first-order approximation to $ T(\fett{F}^{-1}(\fett{x},t),t)$.}

In \cite{Catelan:1996hw,Matsubara:2007wj} the same setting was used \emph{implicitly} in order to derive the longitudinal solution Eq.\,(\ref{finish4thText}) up to the third order in LPT.

%%%%%%%%%%%%%%%%%%%%%%%%%%%%%%%%%%%%%%%%%%%%%%%%%%%%%%%%%%%%%%%%%%%%%%%%
\subsection{Exact relationship between the density contrast and the displacement field}

In SPT, the series of the Fourier transform of the density contrast can be formally written as
\be
  \tilde{\delta}(\fett{k},t) = \sum_{n=1}^{\infty} \tilde{\delta}^{(n)}(\fett{k},t) \,.
\ee
Solving the Eulerian equations of motion order by order in SPT, the resulting $n$th-order density contrast in Fourier space reads:
\begin{eqnarray} \label{SPT}
  \tilde{\delta}^{(n)}(\fett{k},t) = D^n(t) \int &&\frac{\dd^3 p_1 \cdots \dd^3 p_n}{(2\pi)^{3n}}  \, (2\pi)^3 \delta_D^{(3)}(\fett{p}_{1\cdots n} - \fett{k})  \nonumber \\
 &&\ws \hspace{1cm} \times  \, F_n^{(s)}(\fett{p}_1, \ldots, \fett{p}_n) \, \tilde{\delta}^{(1)} (\fett{p}_1) \cdots \tilde{\delta}^{(1)} (\fett{p}_n) \,,
\end{eqnarray}
where the $F_n^{(s)}$'s are the symmetrised SPT kernels given by the well-known recursion relation \cite{Goroff:1986ep,Bernardeau:2001qr}, and are explicitly given in \cite{RampfWong:2012} for $n \leq 4$. 

In order to find the link between the above series and the displacement field, we use again mass conservation, i.e., Eq.\,(\ref{deltajf}).
In Fourier space this is\footnote{The expression $\tilde{\delta}(\fett{k},t) = \int \dd^3 q \exp\{\ii \fett{k} \cdot \fett{q}+ \ii \fett{k} \cdot \fett{\Psi} \} [1- J^F]$ is identical to Eq.\,(\ref{deltaNew}), and leads to a somewhat different expansion but equivalent results.}
\be \label{deltaNew}
 \tilde{\delta}(\fett{k},t) \equiv \int \dd^3 x \,e^{\ii \fett{k}\cdot \fett{x}} \delta(\fett{x},t) 
  =  \int \dd^3 q \, e^{\ii \fett{k}\cdot\fett{q}} \left( e^{\ii \fett{k}\cdot\fett{\Psi}(\fett{q},t)} -1  \right) \,.
\ee
To get the last expression we used $\dd^3 x = |\partial \fett{x} / \partial \fett{q}|\, \dd^3 q$, with $|\partial \fett{x} / \partial \fett{q}| \equiv J^F$.
Taylor expanding the above equation and explicitly using the LPT series, i.e., Eqs.\,(\ref{FTseries}-\ref{finish4thTranseText}), we
can sum up all specific contributions and associate them to their respective $n$th-order density contrasts.
The result is then
\begin{eqnarray} \label{deltaLag}
  \tilde{\delta}^{(n)}(\fett{k},t) &=& D^n \int \frac{\dd^3 p_1 \cdots \dd^3 p_n}{(2\pi)^{3n}}   \, (2\pi)^3 \delta_D^{(3)}(\fett{p}_{1\cdots n} - \fett{k}) \nonumber \\
  &&\ws \hspace{3cm} \times   X_{n}^{(s)}(\fett{k};\fett{p}_1, \ldots, \fett{p}_n) \, \tilde{\delta}^{(1)} (\fett{p}_1) \cdots \tilde{\delta}^{(1)} (\fett{p}_n) \,,
\end{eqnarray}
where we have defined the symmetrised kernels $X_{n}^{(s)}$'s, which can be found in appendix \ref{appX} for $n\leq 4$. The Dirac-delta in Eq.\,(\ref{deltaLag}) fixes $\fett{k} = \fett{p}_{1\cdots n}$ and, as a result,
\be
  X_n^{(s)}(\fett{k};\fett{p}_1,\ldots ,\fett{p}_n)\Big|_{\fett{k} = \fett{p}_{1\cdots n}} \Big. = F_n^{(s)}(\fett{p}_1,\ldots ,\fett{p}_n) \,,
\ee
valid at least up to the fourth-order in perturbation theory. As an important conclusion, Eq.\,(\ref{deltaNew}) leads to identical expansions either in SPT or LPT.

%%%%%%%%%%%%%%%%%%%%%%%%%%%%%%%%%%%%%%%%%%%%%%%%%%%%%%%%%%%%%%%%%%%
\subsection{Exact relationship between the (peculiar-)velocity divergence and the displacement field}

Similar considerations can be made for the peculiar-velocity divergence.
In Eulerian space the peculiar-velocity is $\fett{u} = \dd \fett{x}/ \dd \tau$, where $\dd \tau = \dd t/a$ is the conformal time, and as before $\fett{x} = \fett{q}+\fett{\Psi}$. 
For an EdS universe the time dependence of the fastest growing mode 
is $\propto a^n$ to $n$th order in perturbation theory, therefore $\dd / \dd \tau E(\tau) \fett{\Psi}^{(2)}(\fett{q}) = 2 {\cal H} E(\tau) \fett{\Psi}^{(2)}(\fett{q})$, and similar for higher orders, 
where ${\cal H}$ is the conformal Hubble parameter.
Furthermore, the divergence of the peculiar-velocity is defined by $\fett{\nabla_x} \cdot \fett{u}(\fett{x},\tau)\equiv \theta(\fett{x},\tau)$. Up to the fourth order in LPT the peculiar-velocity divergence is then
\begin{eqnarray} \label{iniT1}
  \theta(\fett{x},\tau) &=& 
  {\cal H} D\, \fett{\nabla_x} \cdot\fett{\Psi}^{(1)}(\fett{q}) +  2  {\cal H} E\, \fett{\nabla_x}  \cdot \fett{\Psi}^{(2)}(\fett{q}) +  3  {\cal H} F\, \fett{\nabla_x}  \cdot \fett{\Psi}^{(3)}(\fett{q}) \nonumber \\
   &&\qquad +  4  {\cal H} G\, \fett{\nabla_x}  \cdot \fett{\Psi}^{(4)}(\fett{q})   \,.
\end{eqnarray}
To proceed it is useful to transform the divergences into Lagrangian coordinates, i.e., 
$\partial/ \partial \fett{x} = |\partial \fett{x}/\partial \fett{q}|^{-1} \partial / \partial \fett{q}$.
Then, Eq.\,(\ref{iniT1}) translates to
\begin{eqnarray} \label{jtheta1}
  \frac{J^F \theta(\fett{x},\tau)}{ {\cal H}} = J^F [{J}_{ij}^F]^{-1} \left( D \Psi_{j,i}^{(1)}(\fett{q})+ 2E \Psi_{j,i}^{(2)}(\fett{q})+ 3F \Psi_{j,i}^{(3)}(\fett{q})+ 4 G \Psi_{j,i}^{(4)}(\fett{q})  \right) \,,
\end{eqnarray}
where $[{J}_{ij}^F]^{-1} = 1/ (2J^F)\, \varepsilon_{ilm} \varepsilon_{jpq} {J}_{pl}^F J_{qm}^F$, with ${J}_{ij}^F = \delta_{ij}+\Psi_{i,j}$. As before commas denote derivatives with respect to Lagrangian coordinates $\fett{q}$.
The Fourier transform of $\theta(\fett{x},\tau)/ {\cal H}$ is
\be \label{FTdiv}
  \frac{\tilde\theta(\fett{k},\tau)}{ {\cal H}(\tau)} = \int \dd^3q  \,J^F\,e^{\ii \fett{k}\cdot \fett{q}+\ii \fett{k}\cdot \fett{\Psi}(\fett{q},\tau)} \frac{\theta(\fett{x},\tau)}{{\cal H}(\tau)}  \,,
\ee
where we have used $\dd^3 x = |\partial \fett{x} / \partial \fett{q}|\, \dd^3 q$, 
and $J^F \theta(\fett{x},\tau)/{\cal H}(\tau)$ is given by Eq.\,(\ref{jtheta1}). We seek a perturbative solution of the form
\be
\tilde\theta(\fett{k},\tau)= \sum_{n=1}^{\infty} \tilde\theta^{(n)}(\fett{k},\tau) \,.
\ee
Taylor expanding Eq.\,(\ref{FTdiv}) and using the LPT results, i.e., Eqs.\,(\ref{finish4thText}, \ref{finish4thTranseText}), we obtain for the $n$th-order peculiar-velocity divergence:
\begin{eqnarray} \label{finalY}
  \tilde\theta^{(n)}(\fett{k},\tau) = - {\cal H} D^n \int &&\frac{\dd^3 p_1 \cdots \dd^3 p_n}{(2\pi)^{3n}} \, (2\pi)^3 \delta_D^{(3)}(\fett{p}_{1\cdots n} -\fett{k} ) \, \nonumber \\
 &&\ws \hspace{1cm} \times \,  Y_n^{(s)} (\fett{k}; \fett{p}_1, \ldots, \fett{p}_n) \, \tilde{\delta}^{(1)} (\fett{p}_1) \cdots \tilde{\delta}^{(1)} (\fett{p}_n) \,,
\end{eqnarray}
where the symmetrised kernels $Y_n^{(s)}$ up to the fourth order are given in appendix \ref{appX}.
Then similar to the above we find
\be
   Y_n^{(s)}(\fett{k};\fett{p}_1,\ldots ,\fett{p}_n)\Big|_{\fett{k} = \fett{p}_{1\cdots n}} \Big. = G_n^{(s)}(\fett{p}_1,\ldots ,\fett{p}_n) \,,
\ee
with $G_n^{(s)}$ given by SPT recursion relation. Thus, the prediction of the peculiar-velocity divergence agrees with its counterpart in SPT, which is:
\begin{eqnarray} \label{SPTdiv}
  \tilde{\theta}^{(n)}(\fett{k},t) = D^n(t) \int &&\frac{\dd^3 p_1 \cdots \dd^3 p_n}{(2\pi)^{3n}}  \, (2\pi)^3 \delta_D^{(3)}(\fett{p}_{1\cdots n} - \fett{k})  \nonumber \\
 &&\ws \hspace{1cm} \times  \, G_n^{(s)}(\fett{p}_1, \ldots, \fett{p}_n) \, \tilde{\delta}^{(1)} (\fett{p}_1) \cdots \tilde{\delta}^{(1)} (\fett{p}_n) \,.
\end{eqnarray}
The SPT kernels $G_n^{(s)}$ up to the fourth order are given in \cite{RampfWong:2012} as well.

%%%%%%%%%%%%%%%%%%%%%%%%%%%%%%%%%%%%%%%%%%%%%%%%%%%%%%%%%%%%%%%%%%%
\section{Discussion and conclusions}\label{sec:discussion}

We have calculated solutions of the Lagrangian perturbation theory (LPT) for an irrotational fluid up to the fourth order. The derivation is shown in two separate approaches, and we call them the full-system- and the peculiar-system approach. They are solved for two sets of `Zel'dovich-type' initial conditions: for the full-system approach we use the slaved initial conditions, where the initial parallelity between the peculiar-velocity and the peculiar-acceleration is assumed; for the peculiar-system approach we use the inertial initial conditions, where the initial peculiar-acceleration is set to zero. Both scenarios are appropriate for studying the large-scale structure.

Solutions up to the fourth order have been found for the first time in \cite{vanselow} (in the full-system approach), whereas one of the current authors independently derived the solutions in the peculiar-system approach.

In section \ref{sec:realisation} we transform the $n$th-order solution into Fourier space, see Eqs.\,(\ref{finish4thText}-\ref{finish4thTranseText}). These solutions contain only the fastest growing mode and consist of integrals over $n$ linear density contrasts.
It is important to note that these results are derived in the so-called \emph{initial position limit} (IPL), in which it is assumed that the aforementioned linear density contrast $\delta^{(1)}(\fett{x})$ is evaluated in the vicinity of the initial Lagrangian position, i.e., in the limit $\fett{x} \approx \fett{q}$. Stated in another way, since the displacement field is small in the Zel'dovich approximation, we can evaluate the results in this limit, and the resulting Poisson equation for the displacement potential assumes therefore the linearised form $\Delta_{\fett{q}} \phi^{(1)}(\fett{q}) \approx  -\delta^{(1)}(\fett{q})$. Working in the IPL requires a small displacement, the starting assumption of the LPT series, but it nevertheless implies that we neglect the inherent non-linearity of the Lagrangian approach in this step. Strictly speaking, only the linear displacement potential and the linear displacement field are affected by this approximation, but due to the reason that the $n$th order displacement field is dependent on $n$ displacement potentials, this approximation carries over to any order. This procedure is implicitly assumed in the current literature when working in Fourier space.

Then, we express the Eulerian dynamical variables, i.e., the density contrast
and the peculiar-velocity divergence in terms of the gravitational induced displacement field $\fett{\Psi}$: the predictions in Fourier space yield identical expressions for the LPT and SPT series at each order in perturbation theory in the IPL.
 However, this does \emph{not} imply, that the convergence of the series in SPT and LPT is identical, rather the above subclasses of the general Lagrangian solution can be used to mimic the SPT series, as long as one restricts to the IPL. 
 Additionally, the IPL suggests that there is \emph{no proper distinction} between the Fourier transform on Eulerian space and on Lagrangian space. Relaxing the IPL would thus result in SPT and LPT solutions in separated Fourier spaces, whereas these spaces are connected via a non-linear transformation. 
This scenario may lead to an even better approximation of the Eulerian variables in terms of LPT. 
However, this issue is beyond the scope of this paper and we leave it as a future project.

%%%%%%%%%%%%%%%%%%%%%%%%%%%%%%%%%%%%%%%%%%%%%%%%%%%%%%%%%%%%%%%%%%
\section*{\sl Acknowledgements:}
{\sl CR would like to thank Yvonne\,Y.\,Y.\,Wong for many fruitful discussions and comments on the manuscript. 
This work was supported by ``F\'ed\'eration de Physique Andr\'e--Marie Amp\`ere, Lyon'', and was conducted within the ``Lyon Institute of Origins'' under grant
ANR-10-LABX-66.}

%%%%%%%%%%%%%%%%%%%%%%%%%%%%%%%%%%%%%%%%%%%%%%%%%%%%%%%%%%%%%%%%%
%%%%%%%%%%%%%%%%%%%%%%%%%%%%%%%%%%%%%%%%%%%%%%%%%%%%%%%%%%%%%%%%%
\appendix

\let\cleardoublepage\clearpage

\section{Preparing the solutions for practical realisation}\label{app:Prep}

Our goal in this appendix is to derive Eqs.\,(\ref{finish4thText}-\ref{finish4thTranseText}). To do so, we first have to transform our spatial $n$th-order solutions, $\fett{\nabla_q} \cdot \fett{\Psi}^{(n)}(\fett{q}) \equiv \Delta_{\fett{q}} \phi^{(n)}(\fett{q})$ and $\fett{\nabla_q} \times \fett{T}^{(n)}$, into Fourier space. Then, in appendix \ref{appTime} and  \ref{app:Res}, we obtain the full $n$th-order solutions $\tilde{\fett{\Psi}}^{(n)}(\fett{k},t)$ and $\tilde{\fett{T}}^{(n)}(\fett{k},t)$ by multiplying the fastest growing mode solution with its spatial $n$th-order solution.

\subsection{Fourier analysis}\label{app:fourier}

In this appendix we prepare the Lagrangian solutions for their use in Fourier space.
Let us start with our Fourier convention,
\be
 \tilde{\fett{A}}(\fett{k}) = \int \dd^3 x \, e^{\ii \fett{k}\cdot \fett{x}} \fett{A}(\fett{x}) 
 \,,\qquad  \fett{A}(\fett{x}) = \int \frac{\dd^3 k}{(2\pi)^3} \, e^{-\ii \fett{k}\cdot \fett{x}} \tilde{\fett{A}}(\fett{k}) \,.
\ee
As mentioned earlier, the $n$th-order longitudinal perturbation $\fett{\Psi}^{(n)}(\fett{q})$ can be written in terms
of a potential $\fett{\nabla_q} \phi^{(n)}(\fett{q})$. 
In Fourier space this means  
\be \label{long}
-\ii\, p_i \tilde{\phi}^{(n)}(\fett{p}) = \tilde{\Psi}_i^{(n)}(\fett{p}) \,.
\ee
We shall first focus on the \emph{net} longitudinal contributions. 
The spatial part of the $n$th-order perturbation, e.g. Eqs.\,(\ref{2ndSol}, \ref{3rdSol}, \ref{4thSol}), yields:
\begin{eqnarray} \label{ansatz}
 {\tilde{\phi}}^{(n)} (\fett{p}) = - \frac{1}{p^2} \int &&\frac{\dd^3 {p}_1 \cdots \dd^3 p_n}{(2\pi)^{3n}}  
 \, (2\pi)^3 \delta_D^{(3)}(\fett{p}_{1\cdots n} - \fett{p}) \,\nonumber \\ 
  &&\times \kappa_{n}(\fett{p}_1, \ldots,  \fett{p}_n) \,{\tilde{\phi}}^{(1)} (\fett{p}_1)\, 
  \cdots {\tilde{\phi}}^{(1)} (\fett{p}_n) \,.
\end{eqnarray}
\noindent In above equation we employed the shorthand notation 
$\fett{p}_{1\cdots n}= \fett{p}_1 + \fett{p}_2 + \cdots + \fett{p}_n$, and
the kernels $\kappa^{(n)}$ are given by:
\begin{eqnarray} \label{kernels}
 \kappa_{2} (\fett{p}_1, \fett{p}_2) &=& \textcolor{white}{-}\frac{1}{2} \left[ p_1^2p_2^2 
  - (\fett{p}_1 \cdot \fett{p}_2)^2  \right] \,, \nonumber \\
 \kappa_{3a} (\fett{p}_1, \fett{p}_2, \fett{p}_3) &=& 
   -\frac{1}{6} \, \varepsilon_{ikl}  \,p_{1i} p_{2k} p_{3l} \,
  \varepsilon_{jmn} \, p_{1j} p_{2m} p_{3n}  \,, \nonumber \\
 \kappa_{3b} (\fett{p}_1, \fett{p}_2, \fett{p}_3) &=& 
 -\kappa_{2} (\fett{p}_2, \fett{p}_3) \frac{ \kappa_{2} (\fett{p}_1, \fett{p}_{23})}{p_{23}^2} 
 \,,  \nonumber \\  
 \kappa_{4a} (\fett{p}_1, \fett{p}_2, \fett{p}_3, \fett{p}_4) &=& 
 \textcolor{white}{-}\kappa_{2} (\fett{p}_1, \fett{p}_3) \,\kappa_{2} (\fett{p}_2, \fett{p}_4)  \,
 \frac{\kappa_{2} (\fett{p}_{13}, \fett{p}_{24})}{p_{13}^2p_{24}^2 }  \,,  \nonumber \\
 \kappa_{4b} (\fett{p}_1, \fett{p}_2, \fett{p}_3, \fett{p}_4) &=& -6\, \kappa_{3a}(\fett{p}_1,\fett{p}_2,\fett{p}_{34})
  \,  \frac{\kappa_{2} (\fett{p}_3, \fett{p}_4)}{p_{34}^2}
 \,,\\
 \kappa_{4c} (\fett{p}_1, \fett{p}_2, \fett{p}_3, \fett{p}_4) &=&  
  -\kappa_{3a} (\fett{p}_2, \fett{p}_3, \fett{p}_{4})  
 \frac{\kappa_{2} (\fett{p}_1, \fett{p}_{234})}{p_{234}^2} \,, \nonumber\\
 \kappa_{4d} (\fett{p}_1, \fett{p}_2, \fett{p}_3, \fett{p}_4) &=&  
 - \kappa_{3b} (\fett{p}_2, \fett{p}_3, \fett{p}_4)  
 \frac{\kappa_{2} (\fett{p}_1, \fett{p}_{234})}{p_{234}^2} \,,  \nonumber \\
 \kappa_{4e} (\fett{p}_1, \fett{p}_2, \fett{p}_3, \fett{p}_4) &=&  
 -\left[ (\fett{p}_1 \cdot \fett{p}_{2})(\fett{p}_{234} \cdot \fett{p}_{34}) 
 - (\fett{p}_1 \cdot \fett{p}_{34}) (\fett{p}_{234} \cdot \fett{p}_2) \right] \nonumber \\ 
    && \ws \hspace{3.5cm} \times \left( \fett{p}_1 \cdot \fett{p}_{234}  \right) 
  \frac{\fett{p}_2 \cdot \fett{p}_{34}}{p_{234}^2 p_{34}^2} \, \kappa_{2} (\fett{p}_3, \fett{p}_4)  
 \,. \nonumber  
\end{eqnarray}
It is worthwile to make the derivation of $\kappa_{4e}$ explicit. In order to write an expression 
for $ {\tilde{\phi}}^{(4e)}$, we first note that the Fourier transform of the transverse part 
(last expression in Eq.\,(\ref{3rdSol})) can be written as
\begin{eqnarray} \label{transe}
 \ii \left( \fett{p} \times \tilde{\fett{T}}^{(3c)} \right)_i 
 = \,&& \int \frac{\dd^3 {p}_1 \dd^3 {p}_2 \dd^3 {p}_3}{(2\pi)^9} \, (2\pi)^3 \delta_D^{(3)}(\fett{p}_{123} 
 - \fett{p}) \frac{(\fett{p}_1 \cdot \fett{p}_{23})}{p_{23}^2}  
 \,\left( \fett{p}_1 \times \fett{p}_{23} \right)_i \, \nonumber \\ 
  &&\times \,\kappa_{2} (\fett{p}_2, \fett{p}_3) {\tilde{\phi}}^{(1)} (\fett{p}_1)\,{\tilde{\phi}}^{(1)} 
   (\fett{p}_2)\, {\tilde{\phi}}^{(1)} (\fett{p}_3) \,.
\end{eqnarray}
Since $\fett{T}^{(3)}$ is purely transverse, we can write it as a curl of a vector potential, i.e., $T_j^{(3)} = \varepsilon_{jlm} A_{m,l}^{(3)}$.
\noindent Using this, Eq.\,(\ref{transe}) reads:
\begin{eqnarray}
  p^2 \tilde{A}_i^{(3)} &-& p_i \left( \fett{p} \cdot  \tilde{\fett{A}}^{(3)}  \right)= 
- \int \frac{\dd^3 {p}_1 \dd^3 {p}_2 \dd^3 {p}_3}{(2\pi)^9} \, 
 (2\pi)^3 \delta_D^{(3)}(\fett{p}_{123} - \fett{p}) \frac{(\fett{p}_1 \cdot \fett{p}_{23})}{p_{23}^2} 
\,\nonumber \\ && \times \left( \fett{p}_1 \times \fett{p}_{23} \right)_i \,\kappa_{2}
 (\fett{p}_2, \fett{p}_3) {\tilde{\phi}}^{(1)} (\fett{p}_1)\,{\tilde{\phi}}^{(1)} 
(\fett{p}_2)\, {\tilde{\phi}}^{(1)} (\fett{p}_3) \,.
\end{eqnarray}
The second term on the LHS may be removed by an appropriate choice of gauge, however, 
this term will vanish anyway. Using Eq.\,(\ref{4thSol}), we can write: 
\begin{eqnarray} \label{transe2}
  &&{\tilde{\phi}}^{(4e)} (\fett{p}) = - \frac{1}{p^2}  
 \int \frac{\dd^3 p_1 \dd^3 p_2}{(2\pi)^6} \, (2\pi)^3 \,\delta_D^{(3)}(\fett{p}_{12} - \fett{p}) 
 (\fett{p}_1 \cdot \fett{p}_2) p_{1j} \varepsilon_{jlm}
  p_{2l} {\tilde{\phi}}^{(1)} (\fett{p}_1)  \\
 &&\qquad\times \Bigg[ \Big\{-\frac{1}{p_2^2} \int \frac{\dd^3 p_3 \,\dd^3 p_4 \,\dd^3 p_5}{(2\pi)^9} \, 
 (2\pi)^3 \,\delta_D^{(3)}(\fett{p}_{345} - \fett{p}_2) \frac{(\fett{p}_3 \cdot \fett{p}_{45} )}{p_{45}^2}
      \Big. \Bigg. \nonumber \\ &&\Bigg. \Big. \hspace{2cm} 
 \times \left( \fett{p}_3 \times \fett{p}_{45}  \right)_m  \kappa_{2} (\fett{p}_4, \fett{p}_5)  
 {\tilde{\phi}}^{(1)} (\fett{p}_3) \,{\tilde{\phi}}^{(1)} (\fett{p}_4)\, 
 {\tilde{\phi}}^{(1)} (\fett{p}_5)  \Big\} 
 + \frac{p_{2m}}{p_2^2}  (\fett{p}_2 \cdot \tilde{\fett{A}}^{(3)}(\fett{p}_2) ) \Bigg] \,, \nonumber
\end{eqnarray}
which yields $\kappa_{4e}$.

Now we proceed with the derivation of the $n$th-order transverse perturbation. The transversality
requirement of $\fett{T}^{(n)}$ in Fourier space reads:
\be
  \tilde{T}_j^{(n)}(\fett{p}) = -\ii\, \varepsilon_{jlm} \,p_l \tilde{A}_m^{(n)} (\fett{p}) \,.
\ee
Using this and the transverse results in Eqs.\,(\ref{3rdSol},\ref{4thSol}), we can write
\begin{eqnarray} \label{ansatzTranse}
 {\tilde{\fett{T}}}^{(n)} (\fett{p}) = - \frac{1}{p^2} \int &&\frac{\dd^3 {p}_1 \cdots \dd^3 p_n}{(2\pi)^{3n}}  
 \, (2\pi)^3 \delta_D^{(3)}(\fett{p}_{1\cdots n} - \fett{p}) \,\nonumber \\ 
  &&\times \fett{\omega}_{n}(\fett{p}_1, \ldots,  \fett{p}_n) \,{\tilde{\phi}}^{(1)} (\fett{p}_1)\, 
  \cdots {\tilde{\phi}}^{(1)} (\fett{p}_n) \,,
\end{eqnarray}
where we have defined:
\begin{eqnarray}\label{transekernels}
 \fett{\omega}_{3c}(\fett{p}_1,\fett{p}_2,\fett{p}_3) &=&
   \left[ \fett{p}_1 \left( \fett{p}_{123} \cdot \fett{p}_{23} \right) - \fett{p}_{23} \left( \fett{p}_{123} \cdot \fett{p}_1 \right)  \right]  \,\left(\fett{p}_1 \cdot \fett{p}_{23} \right)\, \frac{\kappa_2(\fett{p}_2,\fett{p}_3)}{p_{23}^2} 
    \,, \nonumber \\
 \fett{\omega}_{4f}(\fett{p}_1,\fett{p}_2,\fett{p}_3,\fett{p}_4)  &=& 
   \left[ \fett{p}_1 \left( \fett{p}_{1234} \cdot \fett{p}_{234} \right) - \fett{p}_{234} \left( \fett{p}_{1234} \cdot \fett{p}_1  \right)   \right]\,\left( \fett{p}_1 \cdot \fett{p}_{234} \right) \, \frac{ \kappa_{3a}(\fett{p}_2,\fett{p}_3,\fett{p}_4)}{p_{234}^2}
  \,, \nonumber \\
 \fett{\omega}_{4g}(\fett{p}_1,\fett{p}_2,\fett{p}_3,\fett{p}_4)  &=& 
   \left[ \fett{p}_1 \left( \fett{p}_{1234} \cdot \fett{p}_{234} \right) - \fett{p}_{234} \left( \fett{p}_{1234} \cdot \fett{p}_1  \right)   \right] \, \left( \fett{p}_1 \cdot \fett{p}_{234} \right) \, \frac{\kappa_{3b}(\fett{p}_2,\fett{p}_3,\fett{p}_4)}{p_{234}^2}
   \,, \nonumber  \\
  \fett{\omega}_{4h}(\fett{p}_1,\fett{p}_2,\fett{p}_3,\fett{p}_4)  &=& 
   \left[ \fett{p}_1 \left( \fett{p}_{1234} \cdot \fett{p}_{234} \right) 
     - \fett{p}_{234} \left( \fett{p}_{1234} \cdot \fett{p}_1  \right)   \right] \frac{\fett{p}_1\cdot \omega_{3c}^{(s)}(\fett{p}_2,\fett{p}_3,\fett{p}_4)}{p_{234}^2}\,.  
\end{eqnarray}
It is important to note that the $\fett{\omega}$'s satisfy the condition:
\be
  \label{transverseCondition} \fett{p}_{12\cdots n} \cdot \fett{\omega}_n (\fett{p}_1,\ldots,\fett{p}_n) = 0 \,.
\ee
In the following it is also useful to symmetrise the above kernels. The symmetrisation procedure is \cite{Goroff:1986ep}:
\begin{eqnarray}
  \kappa_n^{(s)}(\fett{p}_1, \ldots, \fett{p}_n) &=& \frac{1}{n!}\sum_{i \in S_n} \kappa_n(\fett{p}_{i(1)}, \ldots, \fett{p}_{i(n)}) \,,  \nonumber \\
  \fett{\omega}_n^{(s)}(\fett{p}_1, \ldots, \fett{p}_n) &=& \frac{1}{n!} \sum_{i \in S_n} \fett{\omega}_n(\fett{p}_{i(1)}, \ldots, \fett{p}_{i(n)}) \,. 
\end{eqnarray}

\subsection{Time-evolution of the fastest growing mode for an EdS universe}\label{appTime}

For practical use (that will become clear below),  
we write the time-evolution of the fastest growing mode in terms of the linear growth function
$D(t)$:
\begin{eqnarray}
  \begin{array}{lllllllll} \label{time}
    D_{\textcolor{white}{a}} &= \textcolor{white}{-}a & \quad & \quad & G_a &= - \frac{51}{539} D^4 \quad & \quad & G_f^T &= -\frac{1}{6} D^4 \\
    E_{\textcolor{white}{a}} &= -\frac{3}{7} D^2 &   &  & G_b &= \textcolor{white}{-}\frac{13}{154} D^4  & & G_g^T &=\textcolor{white}{-}\frac{5}{21} D^4 \\
    F_a &= -\frac{1}{3} D^3   &   & &  G_c &= \textcolor{white}{-}\frac{14}{33}D^4 & & G_h^T &= -\frac{1}{14} D^4\;. \\
    F_b &=  \textcolor{white}{-}\frac{10}{21} D^3 &  & & G_d &= -\frac{20}{33}D^4  &&& \\
    F_c &= -\frac{1}{7}D^3 & &  & G_e &= -\frac{1}{11} D^4 &&&
  \end{array} 
\end{eqnarray}
For general cosmologies, $\Omega_m \neq 1$, $\Omega_\Lambda \neq 0$, one can also find approximate 
solutions of the time-evolution, i.e., for Eqs.\,(\ref{2ndTime}, \ref{3rdTime}, \ref{4thTime}). 
For the first-order up to the third-order solutions
this is often performed by employing fitting factors. However, the impact
of $\Omega_m \neq 1$ on the perturbative kernels is very little and can often be neglected, 
and we assume that this is also the case for the fourth-order solution.

\subsection{Perturbative displacement fields}\label{app:Res}

By using Poisson's equation, the linear displacement potentials in Eqs.\,(\ref{ansatz}, \ref{ansatzTranse}) can be linked to (the spatial part of) the linear density contrast: 
$p^2 \tilde\phi^{(1)}(\fett{p}) \approx \tilde\delta^{(1)}(\fett{p})$. This is the initial position limit.
Then, all we have to do is combine the results from the last two sections, e.g., the Fourier transform of
the second-order displacement fields is $\tilde{\fett{\Psi}}^{(2)}(\fett{k},t) = E(t) \tilde{\fett{\Psi}}^{(2)}(\fett{k})$.
Using Eq.\,(\ref{long}) we finally obtain the (symmetrised) result:
\begin{eqnarray} \label{finish4th}
  {\tilde{\fett{\Psi}}}^{(n)} (\fett{k},t) = -\ii D^n(t) \int &&\frac{\dd^3 p_1 \cdots \dd^3 p_n}{(2\pi)^{3n}} 
  \, (2\pi)^3 \delta_D^{(3)}(\fett{p}_{1\cdots n} -\fett{k} ) \, \nonumber \\
 &&\ws \hspace{1cm} \times \fett{S}^{(n)} (\fett{p}_1, \ldots, \fett{p}_n) 
  \, \tilde{\delta}^{(1)} (\fett{p}_1) \cdots \tilde{\delta}^{(1)} (\fett{p}_n) \,.
\end{eqnarray}
$\fett{S}^{(n)}$ are the longitudinal  perturbative vectors, themselves
dependent on the symmetrised $n$th-order kernels $\kappa_n^{(s)}$ (see Eqs.\,(\ref{kernels})):
\begin{equation} \label{Ls}
 \begin{split}
 \fett{S}^{(1)}(\fett{p}_1) &= \frac{\fett{p}_1}{p_1^2} \,, \\
 \fett{S}^{(2)}(\fett{p}_1, \fett{p}_2) 
  &=  \frac{3}{7} \frac{\fett{p}_{12}}{p_{12}^2} \frac{\kappa_2^{(s)}}{p_1^2 p_2^2}\,, \\
 \fett{S}^{(3)}(\fett{p}_1, \fett{p}_2, \fett{p}_3) &=  \frac{\fett{p}_{123}}{p_{123}^2 \,p_1^2 p_2^2 p_3^2} 
      \left[ \frac{1}{3} \kappa_{3a}^{(s)} -\frac{10}{21} \kappa_{3b}^{(s)}  \right] \,,  \\
 \fett{S}^{(4)}(\fett{p}_1, \fett{p}_2, \fett{p}_3, \fett{p}_4) 
  &= \frac{\fett{p}_{1234}}{p_{1234}^2\, p_1^2 p_2^2 p_3^2 p_4^2} 
    \left[  \frac{51}{539} \kappa_{4a}^{(s)} - \frac{13}{154} \kappa_{4b}^{(s)} 
  \right. \\
 &\ws\hspace{4cm}\left.  - \frac{14}{33} \kappa_{4c}^{(s)} 
 +\frac{20}{33} \kappa_{4d}^{(s)} +\frac{1}{11} \kappa_{4e}^{(s)}  \right] \, ,  
 \end{split}
\end{equation}
where we have suppressed the dependences of 
$\kappa_n^{(s)} \equiv \kappa_n^{(s)} (\fett{p}_1, \ldots, \fett{p}_n)$. 

Computing the transverse part $\tilde{\fett{T}}^{(n)}(\fett{k},t)$ is then completely straightforward.  From Eq.\,(\ref{ansatzTranse}) and Eq.\,(\ref{time}) we obtain
\begin{eqnarray} \label{finish4thTranse}
  {\tilde{\fett{T}}}^{(n)} (\fett{k},t) = -\ii D^n(t) \int &&\frac{\dd^3 p_1 \cdots \dd^3 p_n}{(2\pi)^{3n}} 
  \, (2\pi)^3 \delta_D^{(3)}(\fett{p}_{1\cdots n} -\fett{k} ) \, \nonumber \\
 &&\ws \hspace{1cm} \times \fett{S}_T^{(n)} (\fett{p}_1, \ldots, \fett{p}_n) 
  \, \tilde{\delta}^{(1)} (\fett{p}_1) \cdots \tilde{\delta}^{(1)} (\fett{p}_n) \,,
\end{eqnarray}
and the only non-vanishing $\fett{S}_T^{(n)}$'s are
\begin{equation} \label{LsT}
 \begin{split}
 \fett{S}_T^{(3)}(\fett{p}_1, \fett{p}_2, \fett{p}_3) &=  \frac 1 7\frac{ \fett{\omega}_{3c}^{(s)}}{p_{123}^2 \,p_1^2 p_2^2 p_3^2}  \,,  \\ 
 \fett{S}_T^{(4)}(\fett{p}_1, \fett{p}_2, \fett{p}_3, \fett{p}_4) 
  &= \frac{1}{p_{1234}^2\, p_1^2 p_2^2 p_3^2 p_4^2} 
    \left[  \frac 1 6 \fett{\omega}_{4f}^{(s)} - \frac{5}{21} \fett{\omega}_{4g}^{(s)} +\frac{1}{14} \fett{\omega}_{4h}^{(s)}  \right] \, ,
 \end{split}
\end{equation}
up to the fourth order. The $\fett{\omega}$'s are given in Eqs.\,(\ref{transekernels}).
It is important to note that $\fett{p}_{12\cdots m} \cdot \fett{S}_T^{(n)}(\fett{p}_1,\ldots,\fett{p}_n)$ only  vanishes if $m=n$, due to transverseness.

%%%%%%%%%%%%%%%%%%%%%%%%%%%%%%%%%%%%%%%%%%%%%%%%%%%%%%%%%%%%%%%

\section{Kernels for the relationship between SPT and LPT\label{appX}}

The symmetrised kernels for Eq.\,(\ref{deltaLag}) are
\fleq
\be
  {X}_{1}^{(s)}(\fett{k};\fett{p}_1) =  \fett{k} \cdot \fett{S}^{(1)}(\fett{p}_1)  \,, 
\ee
\be
   {X}_{2}^{(s)}(\fett{k};\fett{p}_1,\fett{p}_2)  =  \fett{k}\cdot \fett{S}^{(2)}(\fett{p}_1,\fett{p}_2) 
  +\frac 1 2 \fett{k} \cdot \fett{S}^{(1)}(\fett{p}_1) \,\fett{k} \cdot \fett{S}^{(1)}(\fett{p}_2)\,,
\ee
\bes 
  \label{thirdX} {X}_{3}^{(s)}(\fett{k};\fett{p}_1,\fett{p}_2,\fett{p}_3) = \fett{k} \cdot 
 \left[ \fett{S}^{(3)}(\fett{p}_1,\fett{p}_2,\fett{p}_3)+ \fett{S}_T^{(3)}(\fett{p}_1,\fett{p}_2,\fett{p}_3)  \right]  
\ees
\bes
    \qquad + \frac 1 6 \fett{k} \cdot \fett{S}^{(1)}(\fett{p}_1) \,\fett{k} \cdot \fett{S}^{(1)}(\fett{p}_2) \,\fett{k} \cdot \fett{S}^{(1)}(\fett{p}_3) 
\ees
\be
    \qquad+  \frac 1 3 \left\{ \fett{k} \cdot \fett{S}^{(1)}(\fett{p}_1)\, \fett{k}\cdot \fett{S}^{(2)}(\fett{p}_2,\fett{p}_3) + \text{two perms.}  \right\} \,,
\ee
\bes
  \label{fourthX} {X}_4^{(s)}(\fett{k};\fett{p}_1,\fett{p}_2,\fett{p}_3,\fett{p}_4) =  \fett{k} \cdot \left[\fett{S}^{(4)}(\fett{p}_1,\fett{p}_2,\fett{p}_3,\fett{p}_4)+\fett{S}_T^{(4)}(\fett{p}_1,\fett{p}_2,\fett{p}_3,\fett{p}_4)  \right] 
\ees
\bes
   \qquad+\frac{1}{24}   \fett{k} \cdot \fett{S}^{(1)}(\fett{p}_1) \,\fett{k} \cdot \fett{S}^{(1)}(\fett{p}_2) \,\fett{k} \cdot \fett{S}^{(1)}(\fett{p}_3) \,\fett{k} \cdot \fett{S}^{(1)}(\fett{p}_4)  
\ees
\bes
 \qquad+\frac 1 3 \Bigg\{\frac 1 2 \Bigg.  \fett{k}\cdot \fett{S}^{(2)}(\fett{p}_1,\fett{p}_2) \,\fett{k}\cdot \fett{S}^{(2)}(\fett{p}_3,\fett{p}_4) +\text{two perms.} \Bigg\} 
\ees
\bes
 \qquad+\frac 1 4 \Bigg\{  \fett{k}\cdot \fett{S}^{(1)}(\fett{p}_1)\,\fett{k} \cdot 
 \left[ \fett{S}^{(3)}(\fett{p}_2,\fett{p}_3,\fett{p}_4) + \fett{S}_T^{(3)}(\fett{p}_2,\fett{p}_3,\fett{p}_4) \right]  \Bigg. +\text{three perms.} \Bigg\} 
\ees
\be
  \qquad+\frac 1 6 \Bigg\{ \frac 1 2 \fett{k}\cdot \fett{S}^{(1)}(\fett{p}_1)\,\fett{k}\cdot 
   \fett{S}^{(1)}(\fett{p}_2)\, \fett{k}\cdot \fett{S}^{(2)}(\fett{p}_3,\fett{p}_4) +\text{five perms.}  \Bigg\}  \,,
\ee
and for Eq.\,(\ref{finalY})
\be
   Y_1^{(s)}(\fett{k};\fett{p}_1) = \fett{k} \cdot \fett{S}^{(1)}(\fett{p}_1)  \,,
\ee
\be
   Y_2^{(s)}(\fett{k};\fett{p}_1,\fett{p}_2)  =  -\frac 4 7 \left[ 1- \frac{(\fett{p}_1 \cdot \fett{p}_2)^2}{p_1^2 p_2^2}\right] + \frac 1 2  \fett{k} \cdot \fett{S}^{(1)}(\fett{p}_1) + \frac 1 2  \fett{k} \cdot \fett{S}^{(1)}(\fett{p}_2)   \,.
\ee
\bes
   Y_3^{(s)}(\fett{k};\fett{p}_1,\fett{p}_2,\fett{p}_3)  = \frac{1}{3}  \Bigg\{ \fett{k}\cdot \fett{S}^{(2)}(\fett{p}_1,\fett{p}_2) 
    +\frac 1 2 \fett{k}\cdot\fett{S}^{(1)}(\fett{p}_1)\, \fett{k}\cdot\fett{S}^{(1)}(\fett{p}_2)  
 \Bigg. 
\ees
\bes
    \qquad- \frac 8 3 \fett{p}_{12}\cdot \fett{S}^{(2)}(\fett{p}_1,\fett{p}_2) \,\fett{k}\cdot \fett{S}^{(1)}(\fett{p}_3)  + \Bigg. \text{two perms.}  \Bigg\}  
\ees
\be
  \qquad + 3 \fett{p}_{123}\cdot \fett{S}^{(3)}  -3\frac{\kappa^{(3a,s)}}{p_1^2 p_2^2 p_3^2} 
   + \frac{18}{7} \frac{\kappa^{(3b,s)}}{p_1^2 p_2^2 p_3^2} \,, 
\ee
\bes
  Y_4^{(s)}(\fett{k};\fett{p}_1,\fett{p}_2,\fett{p}_3,\fett{p}_4)  = - \frac{36}{49} \frac{\kappa^{(4a,s)}}{p_1^2p_2^2p_3^2p_4^2} + \frac  6 7 
  \frac{\kappa^{(4b,s)}}{p_1^2p_2^2p_3^2p_4^2}+ \frac 8 3 \frac{\kappa^{(4c,s)}}{p_1^2p_2^2p_3^2p_4^2} -\frac{80}{21} \frac{\kappa^{(4d,s)}}{p_1^2p_2^2p_3^2p_4^2} 
\ees
\bes
 \qquad  -\frac 4 7 \frac{\kappa^{(4e,s)}}{p_1^2p_2^2p_3^2p_4^2} + 4 \fett{p}_{1234}\cdot \fett{S}^{(4)} +
  \frac 1 4 \Bigg\{ -2 \fett{k} \cdot \fett{S}^{(1)}(\fett{p}_1) \frac{\kappa^{(3a,s)}(\fett{p}_2,\fett{p}_3,\fett{p}_4)}{p_2^2p_3^2p_4^2} \Bigg.
\ees
\bes
  \qquad +  \frac 8 7 \fett{k} \cdot \fett{S}^{(1)}(\fett{p}_1)  \frac{\kappa^{(3b,s)}(\fett{p}_2,\fett{p}_3,\fett{p}_4)}{p_2^2p_3^2p_4^2} + \frac 1 6 \fett{k} \cdot \fett{S}^{(1)}(\fett{p}_1)\, \fett{k} \cdot \fett{S}^{(1)}(\fett{p}_2)\, \fett{k} \cdot \fett{S}^{(1)}(\fett{p}_3)
\ees
\bes
 \qquad + \Bigg. 
  \fett{k} \cdot \left[ \fett{S}^{(3)}(\fett{p}_1,\fett{p}_2,\fett{p}_3)+\fett{S}_T^{(3)}(\fett{p}_1,\fett{p}_2,\fett{p}_3) \right] +  \Bigg. \text{three perms.} \Bigg\} 
\ees
\bes
  \qquad+ \frac 1 6 \Bigg\{-\frac 4 7 \fett{k}\cdot\fett{S}^{(2)}(\fett{p}_1,\fett{p}_2)\left[ 1- \frac{(\fett{p}_3 \cdot \fett{p}_4)^2}{p_3^2p_4^2} \right] -\frac 2 7 \fett{k} \cdot \fett{S}^{(1)}(\fett{p}_1)\, \fett{k} \cdot \fett{S}^{(1)}(\fett{p}_2) \left[ 1- \frac{(\fett{p}_3 \cdot \fett{p}_4)^2}{p_3^2p_4^2} \right] \Bigg.
\ees
\be
  \qquad + \Bigg. \text{five perms.}  \Bigg\} + \frac{1}{12} \Bigg\{ \fett{k} \cdot \fett{S}^{(1)}(\fett{p}_1)\, \fett{k} \cdot \fett{S}^{(2)}(\fett{p}_2,\fett{p}_3)  + \text{eleven perms.}  \Bigg\} \,.
\ee

\bigskip
\noindent{\sl The validity of Eq.\,(\ref{deltaLag}) and Eq.\,(\ref{finalY}) with the kernels given in appendix \ref{appX} has been checked with \texttt{Mathematica}, and the respective code can be obtained upon request: \\\url{rampf@physik.rwth-aachen.de}.}

%%%%%%%%%%%%%%%%%%%%%%%%%%%%%%%%%%%%%%%%%%%%%%%%%%%%%%%%%%%%%%%%%%%%%%%%%%%
\section{Used notation}\label{app:notation}

In Tab.\,\ref{tab:notation} we give an overview of the notation used in this work.

\begin{table}[t]

\hspace*{1.5cm}

 \begin{centering}

   \begin{tabular}{lll} 
     $r_i$                    &  Eulerian coordinate $(i= 1,2,3)$     &  Eq.\,(\ref{conti})  \\
     $t$                      &  cosmic time ($\dd/\dd t = \dot{}\,$) &  Eq.\,(\ref{conti}) \\
     $q_i$                    &  Lagrangian coordinate $(i= 1,2,3)$   &  Eq.\,(\ref{q})  \\
     $\eta$                   &  superconformal time ($\dd/\dd \eta = {}'\,$)           &  Eq.\,(\ref{pois}) \\
     $\al$                    &  prefactor in the peculiar Poisson equation             &  Eq.\,(\ref{pois}) \\
     $\rho(\fett{r},t)$       &  density field                        &  Eq.\,(\ref{conti}) \\
     $\fett{v}(\fett{r},t)$   &  velocity field                       &  Eq.\,(\ref{velocity}) \\
     $\fett{g}(\fett{r},t)$   &  acceleration field                   &  Eq.\,(\ref{velocity}) \\
     $x_i$                    &  comoving distance  $(i= 1,2,3)$      &  Eq.\,(\ref{pois}) \\
     $\fett{u}(\fett{x},\eta)$&  peculiar velocity                    & Eq.\,(\ref{drei}) \\
     $\fett{g}_{\rm pec}(\fett{x},\eta)$   &  rescaled peculiar acceleration field      &  Eq.\,(\ref{LN4}) \\
     $\overline \rho (t)$     &  mean density                         &  Eq.\,(\ref{Euler})  \\
     $\delta(\fett{x},\eta)$  &  density contrast                     &  Eq.\,(\ref{density}) \\
     $\fett{f}(\fett{q},t)$   &  trajectory of the fluid particle     &  Eq.\,(\ref{An1}) \\
     $\fett{F}(\fett{q},\eta)$   & peculiar trajectory of the fluid particle    &  Eq.\,(\ref{An2}) \\
     $J$                      &  Jacobian in the full-system approach &  Eq.\,(\ref{Jacob}) \\
     $J^F$                    &  Jacobian in the peculiar-system approach &  Eq.\,(\ref{LN6}) \\
     $\rho_0$                 &  initial density at $\fett{q}$ and $t_0$             &  Eq.\,(\ref{rho0}) \\
     $\fett{p}(\fett{q},t)$   &  perturbation in the full-system approach            &  Eq.\,(\ref{pert1}) \\
     $p_{ij}^c$               &  cofactor element of the perturbation $\fett{p}$        &  Eq.\,(\ref{LN3fin}) \\
     $\fett{\Psi}(\fett{q},\eta)$    &  perturbation in the peculiar-system approach    & Eq.\,(\ref{pert2}) \\
                              & (also called displacement field) & \\
     $\Psi_{ij}^c$            &  cofactor element of the perturbation $\fett{\Psi}$     & Eq.\,(\ref{LN6fin}) \\
     $\fett{\Psi}^{(n)}(\fett{q})$   &  longitudinal perturbation to the $n$th order ($\equiv \fett{\nabla_q} \phi^{(n)})$   & Eq.\,(\ref{pert1}) \\
     $\fett{T}^{(n)}(\fett{q})$      &  transverse perturbation to the $n$th order ($ \equiv \fett{\nabla_q} \times \fett{A}^{(n)}$)      & Eq.\,(\ref{pert1}) \\
     $\epsilon$                &  small and dimensionless perturbation parameter         & Eq.\,(\ref{pert1}) \\
     $q_n(t)$                  &  growth function of the $n$th order in the perturbation & Eq.\,(\ref{pert1}) \\
                               & (full-system approach) & \\
     $D(\eta), E(\eta), \cdots$&  linear growth function, second order growth function, $\cdots$  & Eq.\,(\ref{pert2}) \\
                               & (peculiar-system approach) & \\
     $\mu_a^{(n)}(\fett{q})$  &  $a$th scalar to the $n$th order      & Eq.\,(\ref{mu}) \\
     ${\cal S}(\fett{q},t_0)$ &  initial potential                    &  Eq.\,(\ref{gpec}) \\
     $\fett{S}^{(n)}$         & perturbative vector arising from longitudinal perturbations & Eq.\,(\ref{Ls}) \\
     $\fett{S}_T^{(n)}$         & perturbative vectors arising from transverse perturbations & Eq.\,(\ref{Ls})
   \end{tabular}

 \end{centering}

\caption{Used notation.}
\label{tab:notation}

\end{table}

%%%%%%%%%%%%%%%%%%%%%%%%%%%%%%%%%%%%%%%%%%%%%%%%%%%%%%%%%%%%%%%%%%%%%%%%%%%

\clearpage

%%%%%%%%%%%%%%%%%%%%%%%%%%%%%%%%%%%%%%%%%%%%%%%%%%%%%%%%%%%%%%%%%%%%%%%%%%%%%

\end{document}